# Mars: new insights and unresolved questions

Hitesh G. Changela[1,2] , Elias Chatzitheodoridis[3,4] , Andre Antunes[5] , David Beaty[6], Kristian Bouw[7], John C. Bridges[8], Klara Anna Capova[9], Charles S. Cockell[10], Catharine A. Conley[11], Ekaterina Dadachova[12], Tiffany D. Dallas[13] , Stefaan de Mey[9], Chuanfei Dong[14] , Alex Ellery[15], Martin Ferus[16] , Bernard Foing[9], Xiaohui Fu[17], Kazuhisa Fujita[18] , Yangtin Lin[1], Sohan Jheeta[4], Leon J. Hicks[8] , Sen Hu[1], Akos Kereszturi[19], Alexandros Krassakis[20], Yang Liu[21], Juergen Oberst[22], Joe Michalski[23], P. M. Ranjith[1], Teresa Rinaldi[24], David Rothery[25] , Hector A. Stavrakakis[3] , Laura Selbmann[26], Rishitosh K. Sinha[27] , Alian Wang[28], Ken Williford[6], Zoltan Vaci[2] , Jorge L. Vago[9], Michael Waltemathe[29]  and John E. Hallsworth[13]

[1]Key Laboratory of Earth & Planetary Physics, Institute of Geology & Geophysics, Chinese Academy of Sciences, Beijing, China; [2]Department of Earth & Planetary Science, University of New Mexico, New Mexico, USA; [3]Department of Geological Sciences, National Technical University of Athens, School of Mining and Metallurgical Engineering, Athens, Greece; [4]Network of Researchers on the Chemical Evolution of Life, Leeds, UK; [5]State Key Laboratory of Lunar and Planetary Sciences, Macau University of Science and Technology (MUST), Macau SAR, China; [6]Mars Program Office, Jet Propulsion Laboratory/California Institute of Technology, Pasadena, California, USA; [7]Creative Division, Notion Theory, Miami, Florida 33131, USA; [8]Space Research Centre, School of Physics and Astronomy, University of Leicester, Leicester LE17RH, UK; [9]Human and Robotic Exploration, European Space Agency (HRE/ESA), European Space Research and Technology Centre (ESTEC), Noordwijk, The Netherlands; [10]School of Physics and Astronomy University of Edinburgh James Clerk Maxwell Building, Peter Guthrie Tait Road, Edinburgh EH9 3FD, UK; [11]NASA Ames Research Center, Mountain View, California 94035, USA; [12]College of Pharmacy and Nutrition, University of Saskatchewan, Canada; [13]Institute for Global Food Security, School of Biological Sciences, Queen's University Belfast, 19 Chlorine Gardens, Belfast BT9 5DL, UK; [14]Department of Astrophysical Sciences and Princeton Plasma Physics Laboratory, Princeton University, USA; [15]Department of Mechanical and Aerospace Engineering, Carleton University, Ottawa, Ontario Canada; [16]J. Heyrovsky Institute of Physical Chemistry, Czech Academy of Sciences, Prague, Czech Republic; [17]Institute of Space Sciences, Shandong University (Weihai), Shandong Province, China; [18]Japanese Aerospace Exploration Agency (JAXA), Tokyo, Japan; [19]Research Centre for Astronomy and Earth Sciences, Budapest, Hungry; [20]Worldwide Business Applied Limited, GRC (Governance, Risk & Compliance), Athens, Greece; [21]National Space Science Centre NSSC, Chinese Academy of Sciences, Beijing, China; [22]DLR Institute of Planetary Research, Berlin, Germany; [23]Hong Kong University, Hong Kong, Beijing, China; [24]Department of Biology and Biotechnology, Sapienza University of Rome, Rome, Italy; [25]School of Physical Sciences, The Open University, Milton Keynes, UK; [26]Department of Ecological and Biological Sciences, University of Tuscia, Viterbo, Italy; [27]Physical Research Laboratory, ISRO, Ahmedabad, India; [28]Department of Earth & Planetary Sciences and McDonnell Center for Space Sciences, Washington University, St Louis, USA and [29]Evangelisch-Theologische Fakultät, Ruhr-Universität Bochum, Bochum, Germany

**Abstract**

Mars exploration motivates the search for extraterrestrial life, the development of space technologies, and the design of human missions and habitations. Here, we seek new insights and pose unresolved questions relating to the natural history of Mars, habitability, robotic and human exploration, planetary protection, and the impacts on human society. Key observations and findings include:

– high escape rates of early Mars' atmosphere, including loss of water, impact present-day habitability;
– putative fossils on Mars will likely be ambiguous biomarkers for life;
– microbial contamination resulting from human habitation is unavoidable; and
– based on Mars' current planetary protection category, robotic payload(s) should characterize the local martian environment for any life-forms prior to human habitation.

Some of the outstanding questions are:

– which interpretation of the hemispheric dichotomy of the planet is correct;
– to what degree did deep-penetrating faults transport subsurface liquids to Mars' surface;
– in what abundance are carbonates formed by atmospheric processes;
– what properties of martian meteorites could be used to constrain their source locations;

- the origin(s) of organic macromolecules;
- was/is Mars inhabited;
- how can missions designed to uncover microbial activity in the subsurface eliminate potential false positives caused by microbial contaminants from Earth;
- how can we ensure that humans and microbes form a stable and benign biosphere; and
- should humans relate to putative extraterrestrial life from a biocentric viewpoint (preservation of all biology), or anthropocentric viewpoint of expanding habitation of space?

Studies of Mars' evolution can shed light on the habitability of extrasolar planets. In addition, Mars exploration can drive future policy developments and confirm (or put into question) the feasibility and/or extent of human habitability of space.

## Introduction

Mars, our closest planetary analogue, once had a more substantial hydrological cycle (Wordsworth, 2016), possibly with oceans (Carr and Head, 2003; Redd 2020; Scheller *et al.*, 2021) and lakes shaping its surface and hosting microbial life (Cabrol and Grin, 1999). Over the Red Planet's history, this water cycle (Jakosky, 2021), dynamo (Mittelholz *et al.*, 2020), and igneous activity either diminished or virtually disappeared. Present-day Mars has active seasonal changes in ice caps (Becerra *et al.*, 2020), permafrost (Wray, 2020) and atmospheric composition (Trainer *et al.*, 2019). Undesired terrestrial microbes (Spry *et al.*, 2017), or extant martian life might co-exist on Mars today (Cabrol, 2021).

We are now using orbiters, landers and rovers to explore these tantalizing possibilities (Smith *et al.*, 2020*a*). The next frontier in space exploration is to return samples from and eventually land humans on Mars. The US National Aeronautics and Space Administration (NASA)'s *Perseverance* rover is paving the way for the first Mars-sample return (Farley *et al.*, 2020); a future mission will also bring back those samples. Added to the existing fleet of rovers and orbiters is the Indian Space Research Organization (ISRO)'s *Mangalyaan* orbiter (Lele, 2014), the United Arab Emirates Mars Mission *Hope* (Sharaf *et al.*, 2020) and the China National Space Administration's (CNSA)'s *Tianwen-1* orbiter with the *Zhurong* rover (Wan *et al.*, 2020). They are resolving the geology of the planet, mostly characterizing the surface and near-surface (to centimetre depths). The European Space Agency (ESA)'s and Roscosmos State Corporation (RSC)'s ExoMars *Rosalind Franklin* rover will explore the martian subsurface with a specialized payload designed for life detection down to depths of metres. The Japanese Aerospace Exploration Agency (JAXA) will return samples from the martian moon Phobos for launch in 2024 with the Martian Moons eXploration (MMX) sample-return mission (Usui *et al.*, 2020). SpaceX has an ambitious programme for human settlement on Mars and the Moon (Bramson *et al.*) with the development of the *Starship* vehicle, bringing together a new era of humans in space.

Identifying habitable environments for both microbes (Cockell, 2021) and humans on Mars is a continuous process (Morgan *et al.*, 2021). An important aspect of the current missions is the increasing internationalization of space research (Baitukayeva and Baitukayeva, 2020), providing an opportunity to unite nations in enabling humanity becoming an interplanetary species. Here, we pose questions on Mars relating to its natural history, the development of space technologies, the design of human missions and habitations, planetary protection policy, and impacts of Mars exploration on human society on Earth.

## Early Mars

### Loss of atmosphere

One of the most-striking differences between ancient and current Mars is that it once had a thicker atmosphere compared to the present day; the Noachian Mars atmosphere was more Earth-like in density. This raises the question of when most of the atmosphere was lost. There are compelling observation-based and theoretical calculations indicating that most of the martian atmosphere escaped into space early in the planet's history (Lammer, 2012; Jakosky *et al.*, 2017), when the intensity of extreme ultraviolet (EUV) and the solar wind flux from the young Sun were much higher than today (Ribas *et al.*, 2005). Moreover, the martian dynamo disappeared ∼4.1 Ga ago (Lillis *et al.*, 2013; Mittelholz *et al.*, 2020), leaving present-day Mars with only weak crustal magnetic fields (Johnson *et al.*, 2020). Our understanding of present-day escape of martian atmosphere, and thus our overview of atmosphere losses throughout Mars' history, has improved greatly with observations from NASA's Mars Atmosphere and Volatile EvolutioN (MAVEN) orbiter in conjunction with detailed theoretical models (Dong *et al.*, 2018; Jakosky et al., 2018).

Atmospheric ion escape rates on Mars significantly varied over time, ranging from ∼$10^{27}$ s$^{-1}$ at ∼4 Ga to ∼$10^{24}$ s$^{-1}$ at the present epoch (Fig. 1). Figure 1 also shows that the photochemical escape rate of hot atomic oxygen in the martian exosphere lies between ∼$10^{26}$ s$^{-1}$ at ∼4 Ga and ∼$10^{25}$ s$^{-1}$ today. These simulations are also consistent with the idea that Mars transitioned from an early warmer and wetter world to a desiccated, planet with a frigid surface and tenuous atmosphere. It is noteworthy that space weather events at early epochs of solar evolution, such as Interplanetary Coronal Mass Ejections (ICMEs), could have catastrophically impacted Mars' atmospheric retention. Recently (8 March 2015), an ICME sideswiped Mars and led to a tenfold enhancement in the atmospheric ion escape rate (Dong *et al.*, 2015), confirming that ICMEs – a highly frequent phenomenon of young stars and hence of the young Sun – could have had a direct effect on the removal of Mars' atmosphere.

The habitability of Mars was drastically affected by the loss of the planet's atmosphere. An atmosphere is necessary to maintain surface liquid water, and to protect putative biota from high-energy particles and radiation. The temporal changes of Mars' atmosphere may shed light on exoplanet habitability. Recent numerical and theoretical studies indicate that both magnetized and unmagnetized planets around M dwarfs (stars much smaller and dimmer than our Sun) might be particularly susceptible to the depletion of ∼1 bar atmosphere over sub-Gyr timescales due to the high stellar radiation and particle fluxes within habitable zones (Dong et al., 2020). Such exoplanets orbiting M dwarfs, as well as those around young solar-type stars (Dong et al., 2019), could have been subjected to high atmospheric escape rates early in their history. It is important to take this time-dependence into account when modelling the habitability of Mars or exoplanets.

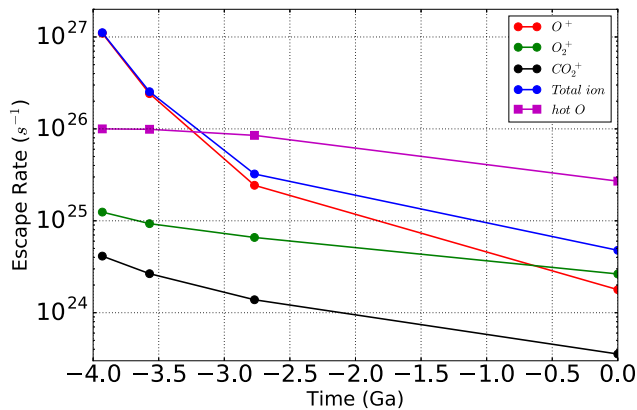

**Fig. 1.** Calculated atmospheric ion and photochemical (hot exospheric oxygen) escape rates over the martian history (under normal solar wind conditions) (Dong et al., 2018).

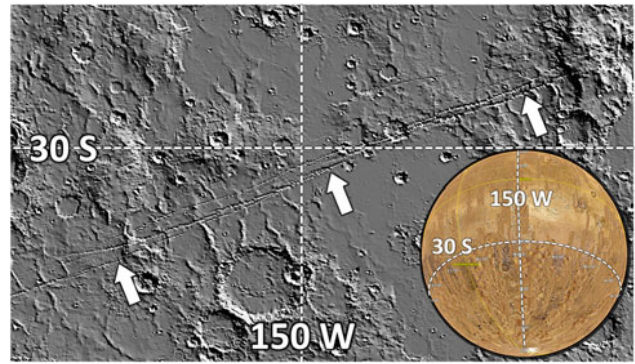

**Fig. 2.** Radial grabens and fractures in the Sirenum Fossae region of southern Tharsis.

### Evidence for and against tectonics on Mars

Early global tectonic-magmatic processes are part of Earth's geological record and have shaped Earth's crust (Stern, 2018). On Mars, radial grabens (Fig. 2) and concentric wrinkle ridges connected to the Tharsis rise (Andrews-Hanna, 2020), and Valles Marineris display properties of fault systems (Mège and Masson, 1996; Baker et al., 2007). A number of smaller tectonic-like features also occur in these regions of Mars. Tectonic processes could account for the alternating magnetized stripes in the southern highlands, possibly due to periodic changes in the magnetic polarity of the martian core, and the gradual spreading at mid-oceanic ridges during crustal formation.

The short length of strike slip faults on Mars may not be related to plate tectonics (Schultz, 1989). Evidence of crustal consumption in the form of subduction zones is also lacking. However, certain faults – abrupt contacts of different rock units and related topographic features – suggest plate boundary-like structural lines (Kidman et al., 2014). Such features would have formed early on in Mars' history. Mars is a planet of modest size planet with limited heat within, leading to relatively rapid rates of cooling and a thick crust. Therefore, a considerable amount of energy must have been required to cause large-scale fracturing of the martian crust. In addition, temperatures in the mantle may not have been high enough to maintain global plate recycling. This could also relate to the depletion of Mars' magnetic dynamo, convention currents, and possible magmatic plumes in its early mantle.

NASA's *InSight* Mars lander mission measured faint seismic activity to shed some light onto the question of Mars' internal structure (Knapmeyer-Endrun and Kawamura, 2020). Recent results suggest a larger, less dense and still-molten core (Stähler et al., 2021). Combined with the relatively thick crust, we should not expect active tectonism. Any large temperature gradients, therefore, must have been created by impacts.

Impact events occurred frequently in Mars' early history. After the cessation of early global impact bombardment, compressive stresses may have occurred for extended periods of time (Watters, 1993), producing wrinkle ridges in several volcanic regions (Schultz, 2000). Moderately small, up to 100 m-long extension-and-compressional features (Plescia and Golombek, 1986) may have shaped these volcanic plains. However, displacements up to 100 km in size have also been observed (Yin, 2012).

Weak plate recirculation does not remobilize as much material from the Mars interior to the surface as occurs on Earth. This produces a low redox gradient; it is the greater redox gradient on Earth that has supported life. However, UV-driven chemistry on the surface of present-day Mars has led to strong redox gradients occurring at depths of centimetres to decametres below the surface. Subsurface measurements, such as those from the ExoMars *Rosalind Franklin* rover (see *Robotic exploration* section), are expected to confirm this.

### What explanation(s) are there for the hemispheric dichotomy?

The lower density of impact craters in Mars' northern hemisphere is coupled with a difference in topographic height between the two hemispheres. The northern hemisphere is lower-lying, and the most-ancient craters there have been buried beneath regolith (and/or rock) material whereas the ancient craters of the southern hemisphere can still be seen (Bouley et al., 2020).

The northern lowlands cover about a third of the planet, and in some places, the southern highlands reach more than 30 degrees north of the equator. The boundary between the northern lowlands and the southern highlands is marked by steep erosional slopes, with apparent traces of ancient shorelines (Sholes et al., 2020), suggesting that the northern lowlands were occupied by an ocean about 3.8–4.1 billion years ago. However, if an ocean existed with a water-depth of several kilometres, the northern lowlands would not have been shielded from impact craters. Why the two hemispheres exhibit dichotomy in their topographic heights is not known (Fig. 3). It has been inferred from studies of the planet's gravitational field that the crust in the north is only about 32 km thick whereas the southern crust is about 58 km thick. The difference in crust thickness should be explained to elucidate the origin of this dichotomy.

It would be tempting to compare Mars' northern lowlands to Earth's thin ocean crust. However, Earth's oceanic crust is produced by seafloor spreading which requires tectonic plates to be moving apart. This type of tectonic process is balanced by convergence elsewhere on the globe (accommodated by subduction, where the edge of one plate descends below the edge of its neighbour). Mars shows no evidence of multiple tectonic plates. In fact, it has been described as a 'one-plate planet'. This difference should be explained to elucidate the origin of this dichotomy.

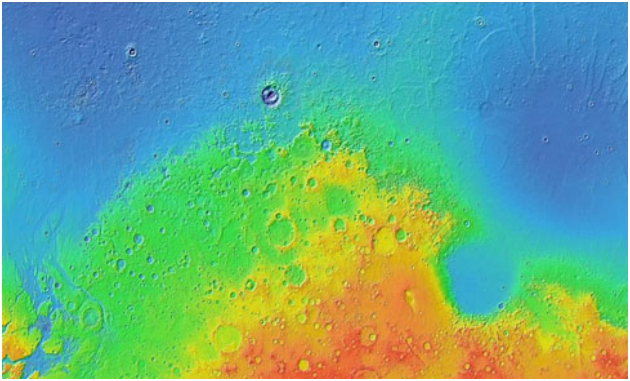

**Fig. 3.** Colour-coded digital elevation map of a latitudinal belt of Mars centred at 24 degrees north. The east-west extent is about 8000 km. Blue-coded elevations are low, whereas green, yellow and red are progressively higher. The total range in this image is about 9 km. The blue, low lying area is north of the dichotomy boundary. It has relatively few superimposed craters, because the surface has relatively young lavas and sediments. The green-yellow-red terrain is south of the dichotomy boundary. This is more ancient and has more craters, despite many of them having been partially erased by erosion. (NASA/JPLK/GSFC/Arizona State University).

There are three distinct theories:

– the low northern hemisphere was excavated by an extremely large impact soon after Mars' formation about 4.5 billion years ago (Andrews-Hanna *et al.*, 2008; Leone *et al.*, 2014);
– a persistent upwelling in the southern hemisphere mantle (Harder and Christensen, 1996) caused partial melting that produced magma which intruded into and erupted onto the planet's earliest crust, resulting in locally enhanced thickness; and
– the southern mantle upwelling was a convection pattern created as a consequence of the northern basin-forming impact (Andrews-Hanna *et al.*, 2008).

The impact event (first theory) would have created a sink for sediments and a site for lava eruptions, both capable of burying the earliest craters that formed on the floor of the giant basin. The second theory looks to processes in Mars' mantle for an explanation. Maybe in a Mars-sized planet (smaller than Earth), convection patterns are simpler, and the sharpness of the dichotomy boundary resulted from erosion rather than reflecting the edge of an impact basin. The third theory integrates the first two theories. Which theory is correct remains unresolved.

### The martian satellites

Phobos and Deimos are the two small natural satellites of Mars with some characteristics similar to asteroids, suggesting that they are objects that were captured in the planet's early history. However, these satellites move in circular orbits that are closely aligned with Mars' equator, which is difficult to explain for captured objects. These orbits are different from that of Earth's Moon, for example, which moves within a more or less ecliptic plane. The orbits of Phobos and Deimos suggest that these satellites formed from large impact ejecta from Mars that then accumulated/coalesced (Bagheri *et al.*, 2021). Alternatively, these satellites may have formed at the same time as proto-Mars itself. The morphology of Phobos is consistent with the idea that the satellite consists of loosely consolidated materials (a so-called rubble pile). Phobos' shape is near ellipsoidal – this could possibly be a relic of the effects of gravitational, rotational and tidal forces during its formation. Such conditions are met at a distance of ∼3.3 Mars radii (our Moon is ∼60 Earth radii away), which is where Phobos probably accreted.

Phobos is orbiting at about 9000 km from the centre of Mars, and therefore is affected by strong tidal forces, resulting in orbital decay. Phobos will likely be disrupted into pieces within a rather short timescale, on the order of a few tens of millions of years. Deimos orbits Mars ∼23 500 km from the centre (∼2.5 times greater than the orbit of Phobos) and is expected to escape from the martian system at some time in the distant future.

The Satellite-Ring Cycle model (Hesselbrock and Minton, 2017) suggests that the moons of Mars were repeatedly created, destroyed, and then recreated. Following a typical tidal disruption of Phobos, a debris ring will form and spread. Most of this material will eventually impact the surface of Mars, but a fraction will migrate into higher orbits. Once beyond the so-called 'Fluid Roche limit', located at about 3.1 Mars radii from the planet's surface, the remaining material will re-accrete to form an offspring moonlet, which will again enter orbital decay. Phobos may be a product of the fourth or fifth generation of the Satellite-Ring Cycle.

## Martian water, alteration minerals and methane

### Lakes and seas on Mars

Life on Earth potentially originated in the ocean (Martin *et al.*, 2008; Osinski *et al.*, 2020), and thrives not only in the modern sea but also in >100 million lakes. It may be in for this reason that there is a widespread interest in ancient lakes and seas on Mars, including the ancient lake of the Jezero Crater that is currently being explored by the *Perseverance* rover. Early reconnaissance of Mars revealed intriguing evidence for many crater lakes (i.e. channels apparently terminating in basins) (Cabrol and Grin, 1999; Horgan *et al.*, 2020), and hinted that the northern hemisphere might have once contained a large ocean (Carr, 1987). Modern assessments, based mostly on laser altimetry, high-resolution visible imaging, hyperspectral compositional remote sensing, and more sophisticated climate models strongly support the idea that many lakes have existed on Mars. However, they have created more questions than answers regarding a northern ocean.

It is widely accepted that several 100s of lakes existed on Mars, mostly during the Noachian (Cabrol and Grin, 2010; Boatwright and Head, 2021). Evidence of open-basin lakes is relatively easy to identify because of outflowing channels which also indicate the maximum height of a base level at some period in time (Fassett and Head, 2008*b*). Underfilled, closed-basin lakes are more difficult to characterize because the lake level, volume and lifespan are unclear. Aside from geomorphological evidence for lakes, compositional remote sensing reveals aqueous alteration minerals within specific lacustrine settings (Ehlmann *et al.*, 2008*a*; Michalski *et al.*, 2019).

On Earth, lakes are geologically transient and on Mars this was also true. Based on estimates of erosion, sediment mobilization, and delta construction it appears that any individual lake might have only existed for $10^4$–$10^6$ years on Mars (Fassett and Head, 2008a; de Villiers et al., 2013). Aqueous mineral assemblages could have formed in similar timescales (Bishop et al., 2018). Placing this in the context of the Noachian period, which lasted ~500 Myr, it is not clear that the existence of lakes implies warm climatic conditions or a substantially thicker atmosphere. It remains a question for debate whether most lakes on Mars formed in relatively short timescales in punctuated climate excursions on an otherwise relatively cold, frozen planet (Wordsworth et al., 2018).

One unusual lake that occurred in Mars' Eridania basin (Fig. 4) in the early Noachian could perhaps more accurately be referred to as a martian sea. This water body would have been kilometres deep, and contained as much water as most of the other lakes on Mars at that time combined. Extremely ancient, it contained thick, deep-water clays of likely hydrothermal origins and coastal evaporites, akin to deposits formed in Earth's oceans today (Michalski et al., 2017).

The case for a vast northern ocean is no stronger now than decades ago, but this hypothesis seems interesting and confusing in equal measure (Saberi, 2020). The vast channels that fed the putative ocean did not all occur at the same time (Warner et al., 2009). It is also not certain that these channels are formed exclusively by the action of water because low-viscosity lava is another possible erosive agent (Leverington, 2011). High-resolution remote-sensing data have been used to search for shorelines as evidence of a northern ocean, but this search has thus-far proved inconclusive. Currently, there is no definitive evidence to indicate that an ancient northern ocean once existed. One problematic issue is that, based on estimates, the amount of water present was insufficient to form a northern ocean, even taking into account the high rates of water loss to space (Carr and Head, 2015).

### Carbonates on Mars

Mars' surface morphology was likely shaped by a warm and wet past (Carr, 1996), though how warm, how wet, and how intermittent these conditions were remains uncertain. Water on the Red Planet would remain in the liquid phase at temperatures maintained by greenhouse gases, mitigating the low intensity of solar radiation on Mars' surface (20–30% lower than that on Earth's surface). It is uncertain whether Mars recycles crustal material (e.g. by plate tectonics) so surface materials ought to remain intact over time. Therefore, the relatively high $CO_2$ pressure ($P_{CO2}$) in the atmosphere of ancient Mars should be evidenced by isotopic and mineralogical substances in the crust, such as carbonates. Geochemical modelling of evaporating mineral sequences in aqueous environments on early Mars, such as a closed basin lake, would likely indicate the precipitation of various carbonates. These might include siderite ($FeCO_3$), calcite ($CaCO_3$), hydromagnesite ($Mg_5(CO_3)_4(OH)_2·4H_2O$), and magnesite ($MgCO_3$) (Catling, 1999). Carbonates on present-day Mars could in this way provide insights into early Mars.

Carbonate-rich outcrops formed through aqueous chemistry have only been detected at a few sites on Mars; by orbital remote sensing and landed missions. These sites include the Nili Fossae grabens that are associated with phyllosilicate, olivine-rich deposits (Ehlmann et al., 2008b) and the Comanche outcrop that is located within the Gusev crater (16–34 wt.%) (Morris et al., 2010). The Jezero Crater also contains carbonate coexisting with olivine and/or Fe/Mg-smectite-bearing outcrops (Goudge et al., 2017; Horgan et al., 2020).

The scarcity of carbonate outcrops on Mars is currently an enigma. However, calculations may have shed some light on this issue. For example, calculations of aqueous equilibria (Fairén et al., 2004) are inconsistent with carbonate formation in an oceanic environment with pH values <6.2, $P_{CO2}$ from 0.4 to 8 bar, and sulphate- and Fe concentrations of 13.5 and 0.8 mM, respectively. Nevertheless, the possibility that many carbonate-rich outcrops occur that are undetected from orbit cannot be excluded. Some sulphates, not identified by orbital remote sensing, were subsequently discovered by landed missions. The *Perseverance*, Tianwen-1 and ExoMars *Rosalind Franklin* rovers may yet reveal martian carbonate outcrops not identified by orbiters.

### Carbonate formation by atmospheric processes

Some carbonates of potential atmospheric origin were found by missions to Mars or by martian meteorite studies. For example, calcium carbonate in the soil around the Phoenix landing site (3–5 wt.%) (Boynton et al., 2009) could form by the interaction of atmospheric $CO_2$ with liquid water films on the surfaces of dust particles. Thermal Emission Spectrometer (TES) data analysis of dust from 21 regions of the martian surface measured 2–5 wt.% carbonates dominated by magnesite (Bandfield et al., 2003), which is also consistent with MiniTES data at the landing site of the *Opportunity* rover where ~5 wt.% carbonate was identified in the regolith (Christensen et al., 2004).

At Gale Crater, *Curiosity* rover's Sample Analysis at Mars (SAM) tool heated sediment and analysed emissions using gas chromatography mass spectrometry (GCMS). $CO_2$ gas was detected at 450–800 °C, a finding consistent with <1 wt.% Fe-, Mg-rich carbonates (Sutter et al., 2019).

Based on the oxygen-isotope study of carbonates in meteorite ALH84001, Farquhar et al. (1998) suggested two oxygen-isotope reservoirs; the atmosphere and the silicate planet. At the time of carbonate growth in this meteorite, the cause of an apparent atmospheric oxygen isotope disequilibrium ($\delta^{17}O$) might be the exchange between the atmospheric $CO_2$ and O ($^1D$) produced by the photodecomposition of ozone. In other words, some carbonate may be generated in the atmosphere by photochemistry.

If carbonate can form from atmospheric $CO_2$ via UVC-induced photochemistry (Farquhar et al., 1998), the energetic electrons resulting from electrostatic discharge (ESD) during martian dust storms (common during the Amazonian) may cause a similar reaction, but through electrochemistry. In a laboratory simulation, $Na_2CO_3$, $NaClO_3$ and $NaClO_4$ were generated after only 1 h in an ESD-Normal Glow Discharge (NGD) process on NaCl under 3 mbar $CO_2$ (Wu et al., 2018). Furthermore, Na-, K-, Al- and Ca carbonates were identified in electrochemical reactions with chlorides under Mars atmospheric conditions (Wang et al., 2020a, 2020b).

### Fe-oxides

Hematite – ferric oxide ($Fe_2O_3$) – is widely distributed across the surface of Mars. Hematite spherules, known as blueberries, have been found in the Meridiani Planum by the *Opportunity* rover. On Earth, hematite can form in a variety of environments and from a variety of parent materials. It is widely accepted that the occurrence of hematite 'blueberries' on Mars indicates that liquid water was once present in Meridiani (Squyres *et al.*, 2004; Golden *et al.*, 2008). However, a clear understanding of the formation of these spherules is still a matter of debate. Various explanations have been investigated, such as the transformation of goethite (FeOOH) and other iron hydroxides (Glotch and Kraft, 2008; Fu *et al.*, 2020), that initially precipitated from aqueous solution and subsequently transformed into hematite. Other theories suggest the role of hydrothermal fluids and seabed water flows depositing the blueberries as observed on Earth (Di Bella *et al.*, 2019), or that the formation of hematite spherules is associated with unique surface conditions on Mars, such as ablation of meteorites (Misra *et al.*, 2014) or freezing aqueous suspensions of hematite nanoparticles (Sexton *et al.*, 2017).

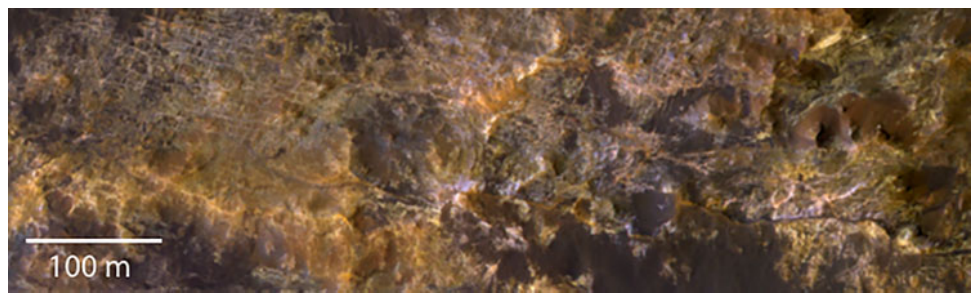

**Fig. 4.** A HiRISE false colour (IRB) image of putative hydrothermal seafloor clays in the deepest parts of Eridania basin. Yellow-brown-green clays are cut by a dense network of veins, and partially covered by younger, dark eolian material.

Hematite deposits at Meridiani Planum were first discovered by the TES on the NASA orbiter *Mars Global Surveyor* (Christensen *et al.*, 2000). Strong absorptions at 315, 461 and 560 $cm^{-1}$ observed by TES are characteristic of coarse, grey hematite (Christensen *et al.*, 2001; Glotch *et al.*, 2004). This finding prompted the selection of Meridiani Planum as the landing site for NASA's Mars Exploration Rovers mission (MER) that includes the *Opportunity* and *Spirit* rovers (Golombek *et al.*, 2003). The images returned by *Opportunity* revealed the abundant accumulations of spherical balls (<0.5 cm diameter). These spherules are ubiquitous on the surface of the martian landscape, shown as blue in the false-coloured images of the rover's panoramic camera (Fig. 5), and dubbed as 'blueberries' (their distribution was considered similar to that of the blueberries in a blueberry muffin) (Squyres *et al.*, 2004). Multiple instruments on *Opportunity* (MiniTES, Pancam, Mossbauer spectrometer and APXS) confirmed that the mineralogy is hematite (Morris *et al.*, 2006). Fragments of these spherules lack an internal structure (Klingelhöfer *et al.*, 2004; Squyres *et al.*, 2004).

The occurrences of these blueberries indicate hydrothermal activity in Meridiani Planum. By contrast, *Spirit* does not find any hematite in Gusev Crater. The Vera Rubin Ridge, also known as the Hematite Ridge, is an erosion-resistant feature within the Gale Crater, associated with a hematite spectral signature according to orbiter-derived data (Fraeman *et al.*, 2013). The *in-situ* X-ray diffraction analysis shows that hematite is present in every Vera Rubin Ridge sample (Rampe *et al.*, 2020) but no blueberries are found within these rocks either.

Fe-oxide concretions found in Utah and Mongolia have been investigated as analogues of martian hematite spherules (Chan *et al.*, 2004; Yoshida *et al.*, 2018). Terrestrial concretions have similar spherical shapes and compositions (Chan *et al.*, 2004; Sefton-Nash and Catling, 2008; Yoshida *et al.*, 2018) though differences between their sizes and mineralogy can occur. Terrestrial concretions (centimetres to metres) are larger than martian blueberries (diameter <6.2 mm) (McLennan *et al.*, 2005; Sefton-Nash and Catling, 2008). Goethite and quartz are major minerals (each present at ≤10 wt%) in terrestrial concretions (Yoshida *et al.*, 2018), whereas martian blueberries are composed of pure hematite.

### Hydrous sulphates

A range of hydrous sulphates occur on Mars (Mg-, Fe-, Al- and Ca sulphates) with Mg sulphate the most abundant. Monohydrated Mg sulphate ($MgSO_4 \cdot H_2O$) (identified as kieserite) and polyhydrated sulphates (interpreted as mostly $MgSO_4 \cdot 4H_2O$ (Wang *et al.*, 2016)) are the most-abundant hydrous sulphates on the martian surface according to remote-sensing studies (Fig. 6). Several-km-thick layers of monohydrated and polyhydrated sulphates have been observed in the Hesperian-aged equatorial regions, such as the *Valles Marineris* interior layered deposits (Ehlmann and Edwards, 2014). Hydrated sulphates have also been found in the Noachian southern highlands (Wray *et al.*, 2009; 2011, Wiseman *et al.*, 2010; Ackiss and Wray, 2014).

Other types of hydrated sulphates, Fe sulphates (e.g. jarosite, hydroxylated ferric sulphates, and szomolnokite) and Al sulphate (alunite) have been identified in highly localized regions of Mars. Large quantities of gypsum occur in the North Polar regions of Mars (Fishbaugh and Head, 2005; Langevin *et al.*, 2005; Fishbaugh *et al.*, 2007; Horgan *et al.*, 2009), while other Ca sulphates have a more localized distribution (Wray *et al.*, 2010; Ackiss and Wray, 2014). The Fe- and Ca sulphates also seem to co-occur with monohydrated and polyhydrated Mg sulphates (Bishop *et al.*, 2009; Lichtenberg *et al.*, 2010; Wray *et al.*, 2011; Weitz *et al.*, 2012; Ackiss and Wray, 2014).

*In-situ* measurements made during surface-exploration missions—Vikings, Pathfinder, Mars Exploration Rovers (MER), Phoenix, and Mars Science Laboratory (MSL)—identified Mg-, Ca- and $Fe^{3+}$ sulphates at all landing sites (Fig. 6). Additional sulphates were found at landing sites (e.g. in Gusev Crater, by the *Spirit* rover) whereas none was detected by orbital remote sensing. In particular, hydrous Mg-, Fe- and Ca sulphates were identified in the trenches and tracks made by *Spirit* rover's wheels (Wang *et al.*, 2006; 2008, Johnson *et al.*, 2007; Arvidson *et al.*, 2010). Furthermore, volatiles (S and Cl) were identified in X-ray amorphous phases in all samples analysed by CheMin (using XRD technology) and SAM (using GC-MS technology) at Gale Crater. Some of these phases might be the alteration products of sulphates.

Remote-sensing observations suggest that most martian hydrated sulphates formed when a relatively large quantity of water was available during the Hesperian period. The chemistry of the hydrous sulphates, being Mg-rich and Fe-, Al-, and Ca-poor, with negligible Na and K, reflects a chemical weathering sequence of basaltic rocks, with higher aqueous dissolution rates for olivine compared to pyroxene, plagioclase and K-feldspar (Hurowitz *et al.*, 2006; McLennan *et al.*, 2005). During hydrothermal episodes, the Ca sulphate veins found in the Endeavor Crater and Gale Crater may have formed (Squyres *et al.*, 2012; Vaniman *et al.*, 2018). Although most hydrated sulphates have been found in the Hesperian terrains, the largest gypsum deposits discovered to date have been found on the Amazonian-aged north polar dunes, which were formed in a unique local environment (Langevin *et al.*, 2005). They have been interpreted to have either mineralized from local rock–water interactions or formed by the erosion of gypsum-bearing materials in the polar ice cap (Horgan *et al.*, 2009).

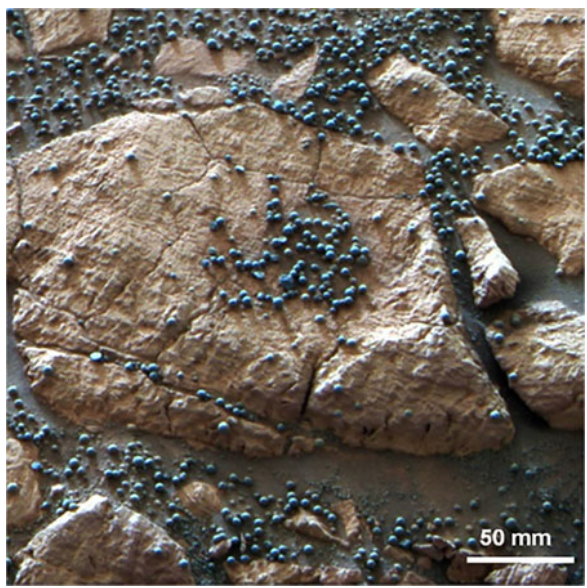

**Fig. 5.** Martian hematite spherules 'blueberries' discovered by the *Opportunity* rover. Image credit: NASA/JPL/USG.

During Mars' more recent past to the present, hydrated sulphates on the surface and in the shallow subsurface have continuously been affected by obliquity changes and the planet's seasonal and diurnal cycles (Laskar *et al.*, 2004). For some sulphates, their degrees of hydration are dependent on the partial water vapour pressures ($P_{H_2O}$) (Kong *et al.*, 2018). Environmental changes on Mars can induce phase transformations of hydrous sulphates, such as those among neutral, acidic and basic $Fe^{3+}$ sulphates (Wang and Ling, 2011). Another potentially important process on present-day Mars is the plasma chemistry induced by martian dust activity. Laboratory simulations have demonstrated the dehydration, amorphization, and oxidation of hydrous sulphates (Wang *et al.*, 2020a).

### Hydrated silicates

The identification of and search for clays on the martian surface has been a key aspect of Mars exploration, in terms of their remote identification by the OMEGA (Observatoire pour la Mineralogie, l'Eau, le Glace e l'Activité) and CRISM (Compact Reconnaissance Imaging Spectrometer for Mars) near-infrared orbiting instruments (Murchie *et al.*, 2009), the selection and characterization of landing sites (Carter *et al.*, 2013), and science activities during landing missions. Phyllosilicates have been detected remotely in layered sediments but are also associated with impact cratering (Turner *et al.*, 2016). Widespread phyllosilicates, particularly within the ancient highlands, include Fe-rich smectite $(NaCa)_{0.5}(Fe^{2+}Mg)_6(SiFe^{3+}Al)_8O_{20}(OH)_4$, chlorite $(FeMg)_5Al(AlSi_3)O_{10}(OH)_8$ and serpentine $(Mg, Fe)_3Si_2O_5(OH)_4$. Fe-rich smectite/saponite (with its characteristic 2.3 μm absorption in CRISM data) was predicted for the Gale Crater central mound – Aeolis Mons (Thomson *et al.*, 2011) before MSL's landing in 2012. Subsequently, CheMin XRD analyses have confirmed the presence of abundant smectite (e.g. Rampe *et al.*, 2017).

### Clay mineralogy and the evolution of the martian crust

A key result in understanding the formation of martian phyllosilicates is evidence that the abundance of clay in Gale Crater fine-grained sediments formed through low-temperature, <50 °C diagenesis – detrital grains such as olivine, plagioclase, etc., reacting with dilute groundwater in buried sediments (McLennan *et al.*, 2014, Bridges *et al.*, 2015). Strong evidence for diagenetic rather than detrital clay origins is given by the textures of sedimentary rocks, e.g. with nodules, veins and ridges (Léveillé *et al.*, 2014). This is a likely origin for much of the clay detected in the ancient highlands. On the basis of the MSL results, we now know that the presence of smectite-dominated clays and an association with layered sediments is an indication of an ancient habitable environment and stable hydrological cycle (Grotzinger *et al.*, 2015). The landing site of Mars2020, Jezero Crater, is partially filled with clay-bearing ancient deltaic deposits (Goudge *et al.*, 2017). It will be important to ascertain if that clay is also predominantly of diagenetic origin, or, alternatively, if detrital clay input was important.

However, the most-closely studied martian clays to date (prior to the MSR mission) are within the nakhlite martian meteorites (Chatzitheodoridis *et al.*, 2014). These have siderite, ferric saponite and ferric serpentine-rich veins (Lee and Chatzitheodoridis, 2016) in their olivine grains and mesostases (Changela and Bridges, 2010; Hicks *et al.*, 2014). The saponite and serpentine formation postdates carbonate mineralization and occurred under near-neutral conditions, and at lower temperatures ∼50 °C (Bridges and Schwenzer, 2012). These meteorites may offer a clue about the origin of phyllosilicates associated with impact craters on Mars. Clays like those preserved in the nakhlites may result from impact-induced hydrothermal activity.

### Methane on Mars

An understanding of methane in the martian atmosphere has been a challenge for the scientific community since the ʹ70s, when the *Mariner* 9 spectrometer IRIS provided the first hints of methane on Mars. The upper concentration limit was estimated at 20 ppb (Mège and Masson, 1996). In 2003, methane detection changed dramatically. Observations made at the Keck Observatory (Keck II Telescope) and the Gemini South Observatory, and the NASA Infrared Telescope Facility (NASA IRTF) in Mauna Kea, Hawaii detected $CH_4$ vibrational bands. In 2004, Krasnopolsky *et al.* (2004), then Formisano *et al.* (2004) reported the ground-based detection of methane in a mixing ratio of 10 ± 3 ppbv during the martian northern summer of 1999 using Fourier Transform Spectroscopy at the Canada–France–Hawaii Telescope.

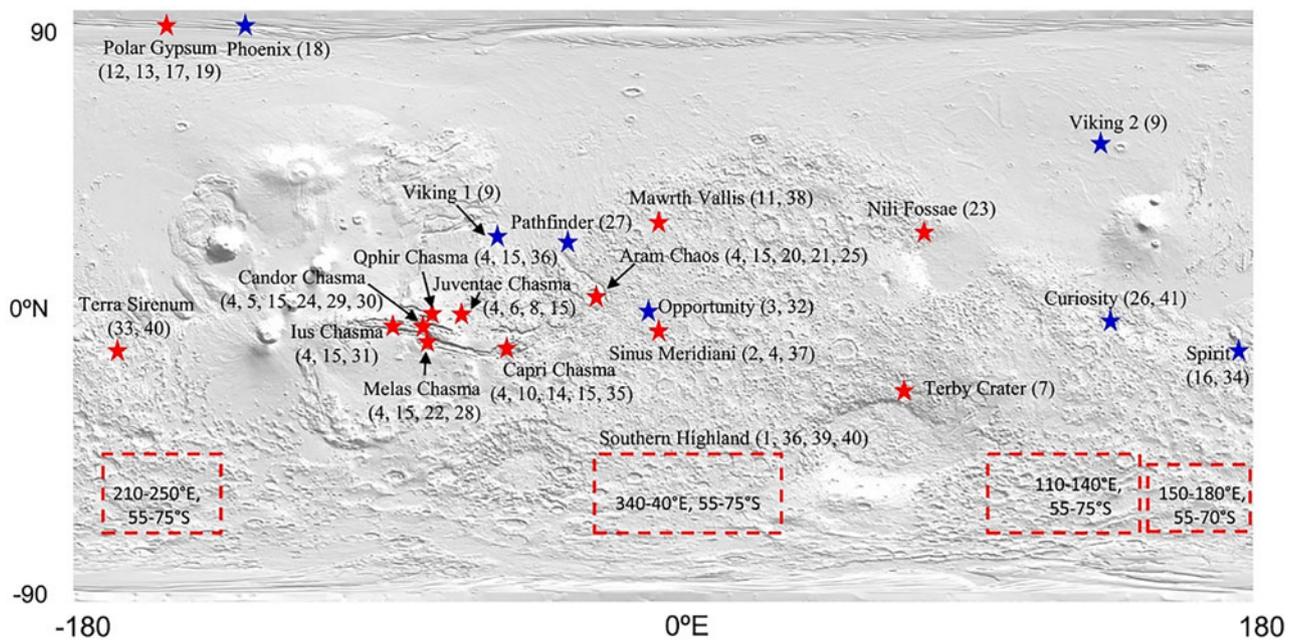

**Fig. 6.** Locations on Mars where hydrous sulphates are found by orbital remote sensing and by surface exploration missions. From (Wang *et al.*, 2016).

Subsequently, Formisano *et al.* (2004) measured methane in the martian atmosphere using the Planetary Fourier Spectrometer (PFS) on board the *Mars Express* spacecraft. The average data from 16 orbits measuring the northern winter hemisphere in 2004 found local variations between 0 and 30 ppbv across the planet. The *Mars Express* data also showed the methane mixing ratio to slowly decrease from the northern spring to the southern summer with an average value of 14 ± 5 ppbv, and to coincide with the water vapour diurnal cycle on Mars (Geminale *et al.*, 2008). The northern mid-summer plume was calculated to contain ∼19 000 metric tons $CH_4$ and interpreted to be from different regions of Mars (Mumma *et al.*, 2009). The TES on-board of *Mars Global Surveyor* over ∼3 martian years (Clancy *et al.*, 2000; Fonti and Marzo, 2010) suggested a seasonal cycle of the methane, as well as inter-annual variations – increases over 32 ppbv with peak values over 60 ppm during the northern summer and autumn, and a decrease during the winter. A subsequent study using the *Mars Express* PFS (Geminale *et al.*, 2011) also documented an increase of methane to a mixing ratio of 34 ppbv over the north polar ice cap during the summer, which cannot be explained by global circulation. Geminale *et al.* (2011) suggested that there could be a methane reservoir associated with the polar ice cap, producing in one year 16 000–24 000 tons $CH_4$, and the methane lifetime could be of the order of 4–6 martian years. Further ground-based observations using the CSHELL (Cryogenic Echelle Spectrograph) and NIRSpec (Near InfraRed Spectrograph) confirmed a far lower maximum limit of methane at 8 ppbv (Zahnle *et al.*, 2011; Krasnopolsky, 2012; Villanueva *et al.*, 2013).

More recently, surface-based measurements of the atmosphere by *Curiosity*, performed for 605 martian sols in the period 2012–2014, measured methane with a background concentration of 0.69 ± 0.25 ppbv (Webster *et al.*, 2015). Occasional spikes were also measured up to 7.2 ±2. 1 ppbv in Gale Crater by the Tunable Laser Spectrometer of SAM. This led to the interpretation that the seasonal variation of methane was 0.24–0.65 ppbv from 2014 to 2017 (Webster *et al.*, 2018). This seasonal variation is also consistent with other atmospheric gas variations in $CO_2$, $H_2O$, $O_2$ and CO (Trainer *et al.*, 2019). However, the magnitude of the seasonal variation in the $CH_4$ background is greater than the other long-lived atmospheric volatiles. This variation may be related to geological and/or meteorological effects, such as losses from photochemistry, sputtering, escape or condensation under any of the conditions reached during the current seasonal cycle (Trainer *et al.*, 2019). It should be noted that the *Curiosity* does not have any instrument that can distinguish between biological and geological sources of methane.

### The origin of methane

On Earth, 95% methane is produced biogenically, by microbial methanogens (Atreya *et al.*, 2007). The process begins with the oxidation of dihydrogen by hydrogenase (equation (1)), followed by the reduction of $CO_2$ (equation (2)) (Thauer, 1998). Methanogens generate energy via the reductive acetyl CoA (Wood–Ljungdahl) pathway, and it is during this process that methane is produced as a by-product (Parkes *et al.*, 2011). Such microorganisms are classified as chemolithoautotrophs because they make their own source of energy by breaking bonds within inorganic molecules (equation (1) and (2)) and can fix $CO_2$.

$$H_2 \leftrightarrow 2e^- + 2H^+ E_0^{'} = 414\,mV \quad (1)$$

$$CO_2 \leftrightarrow 8e^- + 8H^+ \rightarrow 2H_2O + CH_4 \; \Delta G^{0'} = -131\,kJM^{-1} \quad (2)$$

Dihydrogen gas in Equation (1) can also form abiotically as in Equation (3), or degas from erupting volcanoes on Earth.

$$3Fe_2SiO_4 + 2H_2O \rightarrow 2Fe_3O_4 + 3SiO_2 + SH_2 \quad (3)$$

The only way to ascertain the existence of life elsewhere in our Milky Way Galaxy is via spectral signals (Jheeta, 2013). Methane can be generated both abiotically and biotically. It can be made abiotically by serpentinization (Equation (4)). It should be noted that to produce one molecule of methane abiotically requires 26 molecules of water (Equation (4)). Present-day Mars is extensively depleted of water. This may be the reason why the methane plumes that are produced on Mars are only occasional spikes.

$$18Mg_2SiO_4 + 6Fe_2SiO_4 + 26H_2O + CO_2$$
$$\rightarrow 12Mg_3Si_2O_5(OH)_4 + 4Fe_3O_4 + CH_4 \quad (4)$$

Methanogens have two very important attributes, namely they are thought to be the most-primitive organisms on present-day Earth; and they are anaerobic hardy microorganisms able to withstand extreme temperatures and acidities. Such conditions were probably more common on the early Earth. Martian life may be present below the surface, especially if similarly robust microorganisms are harboured within an anaerobic subsurface lake (Westall et al., 2015). This however requires that these microorganisms can utilize dihydrogen in order to produce methane by the reduction of $CO_2$. In anaerobic lakes, there may not be enough dihydrogen for this process which would seep into the atmosphere and then escape to space. For the detection of life on an exoplanet there needs to be a continuous supply of a biogenic gas such as methane maintained at certain levels rather than sporadic tenuous degassing from a planet's surface, as is the case with Mars. Thus, intermittent plumes of methane are more consistent with an abiotic origin.

Recent investigation on the photochemical synthesis of $CH_4$ by the reduction of $CO_2$ over acidic minerals (Shkrob et al., 2012), including a wide range of clays found on both the surface of Mars and in martian meteorites such as Nakhla (Civiš et al., 2017, 2018), can explain not only the formation of methane, but also of perchlorates and chlorinated organic compounds. It can also account for the seasonal variations of methane and carbon monoxide concentrations in the atmosphere (Civiš et al., 2017, 2018). Additional abiotic surface sources involve the decomposition of organics delivered to the martian surface by fragments of comets and asteroids (Fries et al., 2016), volcanic outgassing (Craddock and Greeley, 2009), and slow photocatalytic decomposition of carboxylated molecules in kerogen by photo-Kolbe reactions with Fe(II) oxides in the martian regolith (Shkrob et al., 2010). Finally, the formation of subsurface methane in high pressure-temperature hydrothermal fluids (Welhan, 1988), deep subsurface aquifers (Hu et al., 2016), and the serpentinization of olivine (Oze and Sharma, 2005) cannot be ruled out.

## The habitability of Mars
### Was Mars ever inhabited?

On the question of whether Mars was inhabited, there are two points that we should consider and factor into the experimental search for martian life. We tend to focus closely on the question whether Mars is or was inhabited. However, although it will be difficult to show a definitive lack of life anywhere on Mars, if we eventually explore many potentially habitable environments and find no evidence of life then this would suggest that the planet has been lifeless (Cockell and McMahon, 2019). This finding would be profound. If a planet that was similar to early Earth 4 billion years ago was sterile, yet Earth went on to host about $10^{29}$ microbes (Kallmeyer et al., 2012), this would be more difficult to explain than finding life on another planet that began in a very similar state to Earth. We should remember not to get too focused on a quest to find life on Mars, but rather maintain our interest in testing the hypothesis of life on Mars, either outcome of which would be important.

However, was there life around to take advantage of these habitable conditions? We do not know enough about the origin of life to be able to provide a definitive answer to this (Michalski et al., 2018), nor do we know, despite promising impact-shock experiments, whether panspermia could have successfully occurred between Earth and Mars if Mars was not suitable for an origin of life (Horneck et al., 2008). Further advances in origins of life research on Earth may well elucidate whether Mars had environments suitable for an origin of life, but even if these insights return affirmative answers, that does not demonstrate that this process did occur on Mars. Advances in our understanding of the origin of life and experimental tests of panspermia in shock experiments can never answer definitively the question posed, only provide information that will help us explain what we ultimately observe on the planet itself. We need to go to Mars to unambiguously detect biosignatures.

### What if Mars was never habited?

A habitable, but lifeless Mars would raise important questions. Some people regard such a state as logically impossible – that we cannot define a place as being habitable without the presence of life, but this is incorrect. Microbiologists all the time make agar plates that are habitable to known organisms but are uninhabited (uninhabited habitats). We might find environments on Mars that have all the requirements for known forms of life (such as methanogens), but those environments were never inhabited, either because Mars was always uninhabited, or if it was inhabited, transient habitable environments came and went without being colonized by existing life. Such environments can be demonstrated even on the fecund Earth (Cockell, 2020). These facts do not necessarily bear on the search for biosignatures, but they are important to bear in mind since the paradigm that where there are habitable conditions there is life, which is generally true in most habitable environments on the Earth, may not hold on other planets and clearly would not be true on a planet where there had never been life. Investigating habitable, but uninhabited environments is difficult because one is looking for definitive evidence of the lack of life, which could either be the lack of a biosignature or the presence of some condition that requires a lack of life (such as the presence of some trace element or nutrient that is *always* depleted by life, and whose presence would truly be an anti-biosignature; however, examples of such signatures are difficult to identify). The search for uninhabited habitats raises these observations and experimental complexities.

At the current time, it is not possible to provide a definitive answer to the question, but we do know that the presence of habitable conditions on Mars makes this question a viable and important one to answer.

### Microfossils on Mars

A 'biomarker' on Mars could be in the form of microfossils – fossilized individual or aggregates of microbial cells. A challenge for future exploratory missions would be the unambiguous identification of such features in martian bedrock. In the context of the search for life on Mars, or the earliest records of life on Earth, the term 'microfossil' is generally understood to mean the material remnants of a microbial cell, with recognizably cellular morphology, most commonly preserved as insoluble organic matter, or 'kerogen'. In this sense, there can be no 'microfossil' without fossil morphology.

Can mineralogy, organic chemistry and/or isotopes act as reliable biomarkers without morphology? The specific clarifying examples of isotopes, organic chemistry and mineralogy as potential exceptions might be addressed as follows. In the most-general sense, it could be argued that all three of these phenomena are rooted in form, that is, a spatial ordering of the physical world by life. From this point of view, the 'exceptions' are irrelevant. Certainly, this is true of organic chemistry – purely molecular evidence for biogenicity would occur either as a molecule or set of molecules of sufficient structural complexity that no abiotic mechanism can be established for their synthesis, or a set of molecules with a 'repeating' pattern of different, but related structures.

Minerals are defined in large part by their crystal structure. As such, any biological impact on mineralogy is an expression of form. From another point of view, the intriguing model of 'mineral evolution' advanced by Hazen et al. (2008) maintains that much of the mineral diversity on the modern Earth emerged as at least an indirect result of biology (after biologically mediated accumulation of free oxygen in Earth surface environments). Through this lens, any mineral whose evolution required abundant free oxygen is a biosignature. Would the presence of such a mineral, on its own, be sufficient as definitive proof of an inhabited extraterrestrial planet? This is likely not so. It is certainly possible, though, that a very strong case for the past influence of biology could be made on the basis of a terrestrial planet's mineralogy, but mineralogy alone would likely be insufficient. Biomineralization on Earth produces diverse and diagnostic forms (e.g. in carbonates or iron oxides) that are unknown in abiotic systems, but in these cases, chemical formulae and structure of unit cells are the same as minerals formed abiotically.

Considering isotopic fractionation as an expression of form, there are thresholds of isotopic fractionation in certain Earth systems that are generally accepted as strongly – even conclusively – indicative of biology. Sedimentary organic matter with apparent carbon isotope fractionations (i.e. relative to co-occurring carbonate minerals reasonably expected to record the isotopic composition of inorganic carbon at the time of organic synthesis) >40‰, and sulphide minerals with $\delta^{34}S$ well outside the range between −10 and +10‰ are both strong indicators of biology on Earth. The comparative lack of knowledge about carbon- and sulphur cycles on Mars would make any isotopic observation, on its own, more ambiguous. There is a good reason to believe, however, that a similar abiotic fractionation process acted on Earth and Mars, and large isotopic fractionations between co-occurring oxidized and reduced species containing biologically important elements, expressed in a geologic context consistent with habitability, should be considered as strong potential biosignatures.

### Organics on Mars

Organics discovered from direct analysis of sediments of the martian surface include chlorine (Ming et al., 2014) and sulphur-bearing macromolecular organics (Eigenbrode et al., 2018). In martian meteorites (currently the only martian samples on Earth), nitrogen-bearing organics from the 4-Ga-old martian meteorite, Allan Hills 84 001, are found in carbonates (Koike et al., 2020). Macromolecular reduced carbon in other martian meteorites has also been identified (Steele et al., 2016).

The SAM instrument (Freissinet et al., 2015; Steele et al., 2016) onboard the MSL first measured indigenous organics on Mars above background levels. Analyses of the Sheepbed mudstone in Gale Crater – a sediment with ∼20 wt.% smectite clay interpreted to represent a lacustrine environment – contains chlorine-bearing organic molecules, such as chlorobenzene ($C_6H_5Cl$; 150–300 ppb) and in much smaller amounts, dichloroalkanes (e.g. propane $C_3H_6Cl_2$, ethane $C_2H_4Cl_2$ and butane $C_4H_8Cl_2$) in soil samples from the Cumberland drill hole, interpreted as either indigenous or pyrolyzed by reactions between martian aromatic or aliphatic molecules with oxychlorine phases. Sulphur-bearing organics were also identified by SAM in old mudstone fluviodeltaic sediments from the Murray formation at Pahrump Hills in Gale Crater including thiophenes (i.e. $C_4H_4S$, $C_5H_6S$ and $C_8H_6S$) and thiols ($CH_4S$ and $C_2H_6S$). Aliphatics (chains of one- to five carbons) and aromatic hydrocarbons (benzenes) were among the detected volatiles above the detection limit of the instruments.

Chlorination or sulphurization could be related to abiotic secondary processes on Mars and serve as a means to prevent organics breaking down under intense radiation on the martian surface. Nitrogen-bearing organics in ALH 84001 also have abiotic origins consistent with conditions on Hadean Mars (Koike et al., 2020). More conclusive evidence is expected when the deeper subsurface of Mars is analysed. The Rosalind Franklin rover in Oxia Planum will drill samples 2 m into the subsurface (Martian meteorites and sample return section). The Perseverance rover in Jezero Crater with the Raman and fluorescence SHERLOC spectrometer will search for organics within their mineralogical context. Perseverance is also the first mission caching samples for return back to Earth (Martian meteorites and sample return section). Careful sampling of a variety of geochemical environments and origins will be required. Hydrothermal alteration and vacant areas inside the volume of clays where water activity is higher can not only trap but can also catalyse the synthesis of organics from simpler precursors. Such clays are found in martian meteorites, such as in Nakhla (Chatzitheodoridis et al., 2014) which contain geochemical micro-environment niches where saponite clays form in similarity with, for example, saponite found in basalts near the Mid-Atlantic Ridge of the Atlantis Massif, in which aromatic amino-acids form abiotically (Ménez et al., 2018).

### Disambiguating biogenic and abiotic origins

A biogenic interpretation of organic compounds discovered on Mars based on chemical composition alone is ambiguous; organic chemistry is not exclusive to biology. However, homochirality is a uniquely biological trait. For example, if amino acids were discovered on Mars, a left-handed homochirality would indicate a biological origin based on what we know of life's biochemistry on International Journal of Astrobiology                                                                                                                       11Earth. An organic biomarker could be preserved within a definitive biomorphic structure such as a microfossil (see above). Biogenic organic material also preferentially retains the lighter stable isotope of $^{12}$C. Organic material with any of these properties has yet to be discovered on Mars.

On Earth, organic material is mostly biogenic. An example of an exception is the abiotic organic material delivered to Earth by fragments of asteroids and comets as dust and meteorites. Interplanetary dust would have also delivered an order of magnitude more of organic carbon to Mars than to Earth (Flynn, 1996). The surface of Mars is also peppered by impact craters. Elevated organic distributions around impact craters on Mars could trace their meteoric origins (Frantseva et al., 2018). Higher fractions of stable isotopes heavier than those found in putative biomolecules (e.g. D, $^{13}$C and $^{15}$N) could also imply origins exogenous to Mars. This means that as well as endogenic processes, exogenous delivery should be considered when disentangling abiotic organic evolution on the surface of Mars.

### Putative martian life might have escaped extinction

The endolithic niche is an important target of the search for life elsewhere in the Solar System. Specialized microbes able to exploit this habitat on Earth by developing in the airspaces of rocks are protected from intense solar radiation and desiccation, allowing their population to persist even in the most-extreme terrestrial climates (Walker et al., 2005). Endolithic microorganisms are often the predominant form of life in hot- and cold deserts (Coleine et al., 2021), including the McMurdo Dry Valleys of Antarctica characterized by extreme cold- and arid conditions, along with the intense UV irradiation reaching ground level (Onofri et al., 2004; Bernhard and Stierle, 2020). Therefore, they have value as Mars analogue sites (Friedmann, 1982; Selbmann et al., 2018).

Endolithic life-forms of the Antarctic desert are used as proxies for putative life on planets such as Mars. If putative martian life ever evolved, it may have found a last refuge within the rock pores during the cooling and drying over the early history of the Red Planet (Wierzchos et al., 2011).

### Implications for interplanetary life transfer

Studies on these endolithic communities are pertinent to the concept of lithopanspermia (i.e. the interplanetary transport of microbial passengers inside rocks). Petrographic analyses of martian meteorites indicate that, if during ejection from Mars these rocks experienced shock in the range from 5 to 55 GPa (Artemieva and Ivanov, 2004; Stöffler et al., 2007) and heating in the range from 40 to 350 °C, some of them have never been exposed to temperatures above 100 °C (Weiss et al., 2000; Shuster and Weiss, 2005). Therefore, some fragments of martian crust can be boosted into space whilst experiencing only relatively mild pressure/temperature shocks. Entering into the atmosphere of the recipient planet exposes the meteors to additional fragmentation because speeds, reaching 10–20 km s$^{-1}$, lead to frictional heating melts of the surface. Yet, the short transit time of meteors through the atmosphere, of a few seconds, precludes penetration of heat more than a few millimetres below the surface, limiting the strong heating to <1 mm of depth (meteorite fusion crust) (Fajardo-Cavazos et al., 2005).

Space conditions appear to be the main limit for microbial interplanetary transit; besides, endolithic microorganisms are highly extremotolerant life-forms, able to survive a wide range of injuries, including extremes of temperature, vacuum and high doses of UV and ionizing radiation partially simulating cosmic radiation (up to 60 kGy 60Co) (Selbmann et al., 2018) or Fe ions up to 1000 Gy (Aureli et al., 2020). They also are able to withstand periods as long as 1.5 years in Earth Low Orbit (Onofri et al., 2012; Selbmann et al., 2015; Billi et al., 2019). If shielded and protected by solar irradiation, spores from the (non-endolithic) bacteria *Bacillus subtilis* and *B. pumilus* can survive under Mars-surface simulated conditions (Horneck et al., 2010, 2012; Cortesão et al., 2019). Furthermore, Mars' surface is not necessarily biocidal even for non-sporulating microorganisms (Hallsworth, 2021). Sporulation is a process which appeared early in the development of life on Earth (Tocheva et al., 2016). In addition to aerobic bacilli, present-day endospore formers include anaerobic bacteria, such as the *Clostridium* genus. Endospore-forming bacteria are candidates to survive extreme environments during interplanetary transit (Nicholson and Ricco, 2019). The evidences above show that the planets in our Solar System might not be biologically isolated, because potential micronauts (both prokaryotes and eukaryotes) may have undergone and survived interplanetary transfer.

### Microorganisms resilient to radiation exposure on the Mars surface

Today, ultraviolet- and ionizing radiation through Mars' tenuous atmosphere would seemingly lead to an inhospitable environment for the survival of many kinds of microorganism on the martian surface. However, under some circumstances microbial cells may be able to survive at/near the martian surface (Hallsworth, 2021). Pigmented microorganisms will have better chances of survival because pigments such as melanin are known to be able to absorb/dissipate electromagnetic radiation such as UV and ionizing radiation. The presence of melanin correlates with increased survival and enhanced robustness, resulting in an advantage for melanized organisms over non-melanized ones under harsh conditions. In this regard, melanized fungi are a good example of the survival advantage conferred by melanin. There is a high incidence of melanized fungi in such extreme environments, such as the damaged nuclear reactor at Chernobyl (Zhdanova et al., 1991; Dighton et al., 2008) with 80% of the fungal species recovered from the reactor being melanized. Another example are the rocky deserts of Antarctica (Selbmann et al., 2015). It has also been demonstrated that melanized fungi can survive simulated Mars-like conditions (Onofri et al., 2008) and cosmic radiation while exposed on the Mir Spacecraft (Novikova, 2004). Besides exposure to high doses of ionizing and/or UV radiation, inhospitable environments can deliver additional forms of stress such as salinity, aridity, rapid and extreme temperature fluctuations, as well as little-to-no nutrients. The advantage that melanin presence confers on fungi is based, first of all, on its ability to act as a physical shield, reducing the relative biological effectiveness of ionizing radiation and reducing its potential for damaging living cells. Secondly, melanin protects the organism further by 'scavenging' reactive oxygen species generated by ionizing radiation (reviewed in Malo and Dadachova (2019)). Thirdly, melanin can convert solar radiation to heat. Thus, physical and chemical protective qualities of melanin can result in increased cell survival, and maybe even favour metabolic activity (and even growth) of melanized species under some conditions close to those that can occur on the martian surface. Analyses of the martian atmosphere show that the temperature : water-activity regimes would

Downloaded from https://www.cambridge.org/core. IP address: 2.96.198.120, on 01 Dec 2021 at 15:50:50, subject to the Cambridge Core terms of use, available at https://www.cambridge.org/core/terms.
https://doi.org/10.1017/S1473550421000276

not allow metabolic activity of any known terrestrial microorganisms (Hallsworth *et al.*, 2021).

## Robotic and remote exploration of Mars

### Mars Science Laboratory

The MSL *Curiosity* rover mission achieved a milestone in Mars exploration, landing a car-sized payload on Mars, manoeuvring through Gale Crater, climbing the flank of Aeolis Mons (informally known as Mount Sharp) and performing unprecedented science on the way. *Curiosity*'s traverse path has been determined by both science and safety considerations.

*Curiosity*'s traverse path is determined by science priorities across a variety of spatial and temporal scales. The primary goal for the MSL mission is to explore and quantitatively assess a local region on Mars' surface as a potential habitat for life, past or present. Gale Crater was selected as the landing site for the MSL mission in part because it contains a 5 km tall mound of stratified rocks (known as Aeolis Mons) with intriguing mineral signatures that may record changes in environmental history (c.f. (Milliken *et al.*, 2010; Golombek *et al.*, 2012; Grotzinger *et al.*, 2012)). In order to meet this mission objective within Gale Crater, the traverse path is focused on the long-term goal of ascending the lower flanks of Aeolis Mons, with intermediate traverse decisions determined by focused science campaigns, and daily traverse decisions based on opportunistic science.

*Curiosity* landed on Bradbury Rise, initially mapped as hummocky terrain within the landing ellipse. This area is separated from Aeolis Mons by a large dune field known as the Bagnold Dunes. While the primary scientific goal is to investigate strata at the base of Aeolis Mons, due to traversability constraints the rover had to drive more than 12 km to the southwest to reach a gap in the dunes. The dunes were well-mapped prior to landing, so the general traverse path for the rover to ascend Aeolis Mons was also known since before landing.

While the long-term path has been predetermined, there are intermediate stops along the way based on focused science campaigns. An early example is the exploration of 'Yellowknife Bay', a region just to the east of the landing site that represents the intersection of three distinct geologic units as mapped from orbit. Although the long-term traverse path suggested that driving to the southwest was the way to ascend Aeolis Mons, the team decided to divert from the most efficient path to investigate this area first. This decision paid off. Yellowknife Bay lies at the distal extent of a large alluvial fan that extends from the northern crater rim; once the rover investigated the fine-grained fractured outcrops, we discovered that they represent ancient lake deposits. These lake deposits contain all of the chemical ingredients necessary for life and minerals that would have provided a source of energy for primitive organisms, so this campaign resulted in evidence for the first habitable environment explored by the rover (Grotzinger *et al.*, 2015). Since then, *Curiosity* has explored about a dozen key waypoints for more detailed science campaigns. Some of these campaigns were motivated by orbital observations (e.g. Ehlmann *et al.*, 2017; Rice *et al.*, 2017; Bennett *et al.*, 2019; Fraeman *et al.*, 2020), while others were based on discoveries on the ground (e.g. Banham *et al.*, 2018; Edgar *et al.*, 2018; Stack *et al.*, 2019).

Finally, on the short-term scale, daily traverse decisions are influenced by science through selection of the 'end-of-drive' location. On any given day, the science team works with the engineers to select drive locations that will provide adequate outcrops for future science analyses, or that provide a good perspective on the stratigraphy. The team is keeping an eye on the long-term path, but this is still a mission of exploration, and is ready to respond to findings as they arise.

### Mars 2020

The Mars 2020 mission consisting of the *Perseverance* rover and *Ingenuity* drone landed successfully in Jezero Crater. The mission is dedicated towards geology and astrobiology. *Ingenuity* performed the first ever flight from the surface of another planet.

Jezero is an impact crater ∼45 km in diameter, located at 18.4° N, 77.7°E in the Nili Fossae region, forming a fluvial sedimentary basin (Goudge *et al.*, 2015). The landing site is close to the deltas of rivers that once flooded the crater. Aeolian sediments are also present, laying onto a volcanic floor. Multiple short flooding events composed of a variety of detrital sediments transported from very different provenances are suggested to have deposited the north and western watershed fans, increasing the probability of identifying materials of different origins accumulated around the landing site of *Perseverance*. This could have astrobiological implications, but is also important for understanding climatic processes on Mars such as the regularity of floods, its connection to periodic climatic epochs, and their possible connection to impact events. Apart from the transported and deposited sediments, authogenic clays might be also located and sampled; it is known that clays can record climatic conditions and also trap and preserve organics that are putative biosignatures, either chemical or textural. The close inspection of the samples *in situ* or in laboratories after sample return will enable the search for biosignatures or pseudo-biosignatures. Simulations run in the Mojave Desert have shown that PIXL (Planetary Instrument for X-ray Lithochemistry) can resolve trace element patterns that could be used to select samples for geochronology (Martin *et al.*, 2020). The same applies to the identification of primary mineral phases and volcanic glasses. Resolving secondary mineral phases (i.e. carbonates, clays, sulphates) will be important to elucidate past environmental conditions. Combining all the above instruments will be paramount to identifing undisputed life on Mars.

On board the *Perseverance* rover, there are various optical cameras and microscopes for navigation, analysis and closer characterization. Navigation and remote identification of areas of interest is performed with the Navcam stereo camera, the Mastcam-Z hyperspectral imager and the SuperCam imaging and analytical system (remote Raman and Laser-Induced Breakdown Spectroscopy (LIBS) analysis). Chemical and mineralogical analyses can be performed by all spectroscopy instruments, with the advantage of organic compound detection and mapping with the Scanning Habitable Environments with Raman & Luminescence for Organics & Chemicals (SHERLOC) instrument (Martin *et al.*, 2020). SHERLOC operates with a deep UV laser beam which can be scanned over a small area of a sample, providing mineralogical mapping over a $7 \times 7$ mm area with a spot size of 100 $\mu$m, also captured with the WATSON colour imaging system (Williford *et al.*, 2018). At the same time, SHERLOC acquires fluorescence images, enabling the identification and classification of the composition as well as the mapping of the spatial distribution of organic compounds. The detection sensitivity of organics can go down to $10^{-5}$ to $10^{-6}$ w/w over the entire scanned area. Both instruments are set up to investigate the habitability of the explored area based on the availability of CHNOPS chemistry. Mineral distribution patterns together with organic compounds will enable the identification of putative textural biosignatures.

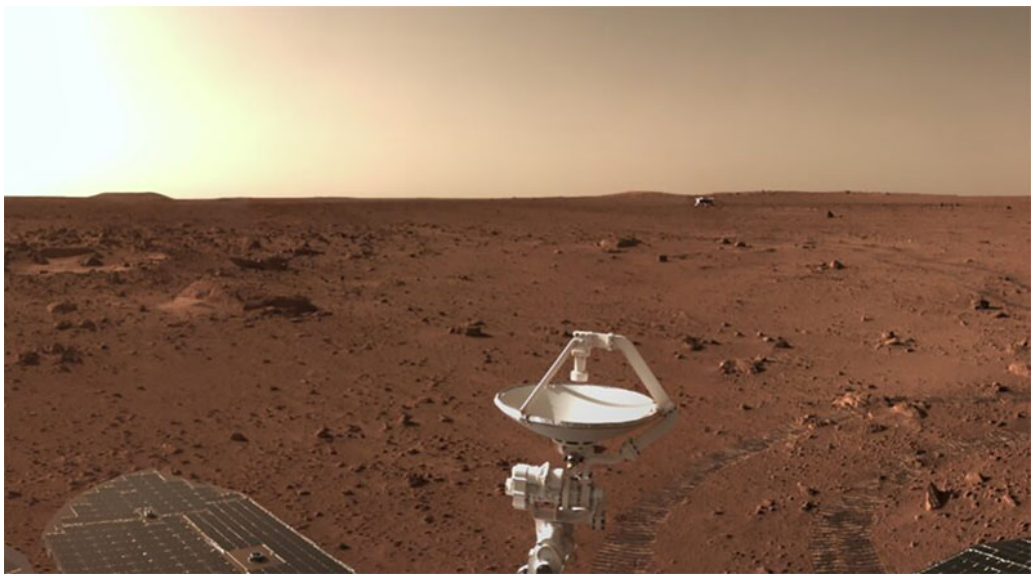

**Fig. 7.** The landscape of Mars at Utopia Planitia taken with the Zhurong rover's Navigation and Topography Cameras.

### Tienwen-1

The CNSA's Tienwen-1 spacecraft planetary mission is a threefold mission, involving an orbiter, a lander and the *Zhurong* rover. The Tienwen-1 lander and *Zhurong* rover landed on Mars on 14 May 2021. *Zhurong* deployed from the lander and touched the martian surface on 22 May 2021 in Utopia Planitia, a plain located within the Utopia impact basin (Fig. 7). The geological environment of Utopia Planitia is being studied, with a focus on volatiles and water in the region.

Utopia Planitia was formed during the Late Hesperian to Mid-Amazonian period, with mainly volcanic rocks and glacial clastic sediments. This region is known from NASA's Viking 2 landers to have permafrost that has shaped the surface by repeated freezing and thawing cycles. Estimating the amount of water in the martian subsurface is of importance for human missions to Mars.

The orbiter was used to select the landing area, relay data back to Earth, and remotely map Mars. The lander was used to carry the rover and commission it on the surface, but also employs a camera with which it took the first panoramic images to identify safe paths for driving the rover. Video footage was also obtained with a remote camera on the surface. It also has a multispectral imager for mineralogical identification and distribution of the surrounding area. *Zhurong* also employs a sub-surface penetrating radar to investigate the 3D stratigraphic history of the area (Wang *et al.*, 2021), and estimate the consistency of the regolith and the depth of the permafrost. This will provide new insight into the ancient water present on Mars and climatic changes during the Red Planet's history. The magnetic field of the area will be measured with a magnetometer, while a LIBS instrument will analyse rocks, seeking very ancient martian rocks. The magnetic field of the planet will be studied remotely from the orbiter with a second magnetometer and a particle analyser. The rover employs instruments for meteorological studies, such as a microphone returning sounds of Mars.

### Mars Orbiter Mission

The ISRO successfully launched the Mars Orbiter Mission (MOM), a probe also known as *Mangalyaan*, aboard the Polar Satellite Launch Vehicle (PSLV)-C25 on 5 November 2013 to Mars. MOM settled into martian orbit on 24 September 2014 and has been observing Mars ever since, far exceeding its 1-year design lifetime. The primary scientific goal of MOM is to explore the morphology, topography and mineralogy of surface features, and analyse the atmosphere using the indigenously developed instruments (Arunan and Satish, 2015). The MOM's payload includes a Mars Colour Camera (MCC) for optical imaging (Arya *et al.*, 2015), a Thermal Infrared Imaging Spectrometer (TIS) for surface temperature estimation and mapping composition and mineralogy (Singh *et al.*, 2015), a Mars Exospheric Neutral Composition Analyser (MENCA) to measure *in-situ* composition of the low-altitude neutral exosphere and its radial distribution (Singh *et al.*, 2015), a Lyman Alpha Photometer (LAP) to determine the deuterium-to-hydrogen abundance ratio of the upper atmosphere (Sridhar *et al.*, 2015), and a Methane Sensor for Mars (MSM) to measure total column of methane in the atmosphere (Mathew *et al.*, 2015).

The large and highly elliptical orbit of the MOM (261 km Perihelion to 78 000 km Apoareion) has enabled the MCC to image the far-side of Mars' moon Deimos for the first time after more than three decades (Mathew *et al.*, 2015). Subsequent analysis of the temporal and on-demand stereo images of the MCC led to the study of haze variability inside Valles Marineris and estimation of the optical depth of the martian atmosphere over the northern and southern walls of the valley (Mishra *et al.*, 2016). Observations of the MCC have also enabled local-scale dust storm tracking over the Lunae Planum region and the Valles Marineris region (Guha *et al.*, 2019).

MENCA observations have led to significant discoveries including *in-situ* compositional measurements for three major constituents of the martian exosphere, i.e. amu 44 ($CO_2$), amu 28 ($N_2$ + CO), and amu 16 (O) corresponding to the local evening hours (Bhardwaj *et al.*, 2015). These measurements are considered critical for setting up the important boundary conditions in the thermal escape models (Bhardwaj *et al.*, 2017). Furthermore, the presence of suprathermal argon atoms (argon-40) has been reported in the martian exosphere using scale height and temperature measurements (Bhardwaj *et al.*, 2015). To establish the boundary conditions for global circulation model calculations, the reflected solar radiance measured by the MSM in two Shortwave Infrared (SWIR) (1.64–1.66 μm) channels has been used to prepare an albedo map of the martian surface (Singh *et al.*, 2017).

After 5 years, MCC continues to capture intriguing views of Mars contributing significantly to the exploration of its surface. There is much more to be leveraged from the MOM datasets that can be fundamental in developing new insight for the surface and atmospheric evolution of Mars. The identification of scientific knowledge gaps and the optimization of suitable payloads are the key to success of the ISRO's continued planetary exploration programme.

### Exomars 2022

The *Rosalind Franklin* rover of the ExoMars 2022 mission (Fig. 8) will land in Oxia Planum (Quantin-Nataf *et al.*, 2021), an ancient location with strong potential for past habitability and preserving physical and chemical biosignatures (as well as abiotic/prebiotic organics). Oxia Planum is situated on the eastern margin of the Chryse basin, along the martian dichotomy border, and at the outlet of the Coogoon Valles system. At present, the coordinates for the nominal touchdown location are 18.159°N, 24.334°W. The approximately 90 × 4 km dispersion ellipse lies in the lower part of a wide basin, where extensive exposures of Fe/Mg-phyllosilicates (>80% of the ellipse surface area) have been detected with OMEGA and CRISM hyperspectral and multispectral data. The Fe/Mg-rich clay detections are associated with early/middle- to late-Noachian layered rocks (with layering thickness ranging from a few metres to <1 m for several tens of metres) (Mandon *et al.*, 2021). They may represent the southwestern expansion (lowest member) of the Mawrth Vallis clay-rich deposits, pointing to a geographically extended aqueous alteration environment, and perhaps an ocean.

### Optimizing life detection with ExoMars

The payload of the ExoMars rover has been put together based on the assumption that it is highly unlikely that lifeforms would exist on the surface of Mars today. Therefore, the focus was placed on the identification of traces of past life, from the time when the surface environment was able to host abundant liquid water, i.e. the Noachian (Vago *et al.*, 2017). With this in mind, ESA decided to concentrate on the detection of physical (textural) and chemical biosignatures. Other possible, but less definitive classes of biosignatures, like isotopic analysis and the investigation of possible metabolites, will not be investigated with the current payload. To maximize the chance of having access to well-preserved biomolecules, the rover is equipped with a drill having a 2-m depth reach. This drill can be used to collect samples from top- and subsurface sedimentary rocks, which are then analysed for mineralogy and organic composition using instruments in the rover's analytical laboratory. The rover exploration strategy is based on a stepwise execution of nested-scale investigations, proceeding from the panoramic visual scale, to the microscopic, and concluding with molecular analyses. The rover's Pasteur payload includes panoramic instruments PanCam (Coates *et al.*, 2017), wide-angle and high-resolution cameras; ISEM (Korablev *et al.*, 2017), an infrared spectrometer; WISDOM (Ciarletti *et al.*, 2017), a ground-penetrating radar; and ADRON (Litvak *et al.*, 2008), a neutron detector; a subsurface drill to acquire samples; contact instruments for studying rocks and collected material (CLUPI (Josset *et al.*, 2017), a close-up imager; and MaMISS (De Sanctis *et al.*, 2017), an infrared spectrometer in the drill head); a Sample Preparation and Distribution System (SPDS); and the analytical laboratory, the latter including MicrOmega (Bibring *et al.*, 2017), a visual and infrared imaging spectrometer; RLS (Rull *et al.*, 2017), a Raman spectrometer; and MOMA (Goesmann *et al.*, 2017), a Laser-Desorption, Thermal-Volatilization, Derivatization, Gas Chromatograph Mass Spectrometer (LD + Der-TV GCMS).

### Martian meteorites and sample return

A few decades ago, it was realized that among meteorites on Earth, about a dozen had textures, young ages, oxygen isotopic ratios, and noble gas compositions consistent with a martian origin. Today there are ∼140 martian meteorites. In particular, noble gas abundances trapped in shocked glass from martian meteorites are in a ∼1:1 ratio with Mars' atmospheric composition.

As well as being trapped from Mars, noble gases can also form in martian meteorites *in situ*. They can form from spallation, radioactive decay and fission reactions called cosmogenic, radiogenic and fissiogenic, respectively. Cosmogenic noble gas components, for instance, help us to trace the ejection time of rocks from the martian surface. Radiogenic noble gas nuclides, such as $^{40}$Ar produced through the decay of $^{40}$K, enable formation ages to be determined. $^{129}$Xe from the decay of $^{129}$I, an extinct radionuclide with a half-life of 15.7 Myr, can unravel atmospheric evolution. Fissiogenic heavy Xe isotopes (e.g. $^{134}$Xe and $^{136}$Xe) in martian meteorites record fission reactions of $^{238}$U and $^{244}$Pu from the martian atmosphere or interior (Smith *et al.*, 2020*b*).

Cosmic ray exposure (CRE) age is the time a meteoroid spent in interplanetary space from ejection to arrival to Earth. Cosmogenic noble gases ($^{3}$He, $^{21}$Ne, $^{38}$Ar, $^{83}$Kr, $^{124}$Xe) produced in the meteoritic material by cosmic ray spallation reactions in specific target elements are measured to determine CRE ages. The ejection time is the sum of CRE age in space and the terrestrial residence time on earth. Terrestrial ages are determined by measuring radionuclides such as $^{14}$C, $^{10}$Be, $^{36}$Cl and $^{41}$Ca. As the terrestrial residency times are negligible compared to the CRE ages, CRE ages directly indicate ejection times. Source-crater pairings of different martian meteorites are established on the basis of CRE age groupings and their petrologic affinities (Gladman, 1997). The martian meteorites group into 9–10 different CRE ages. Thus, there were around 9–10 different ejection events on Mars, which produced the current suite of martian meteorites that arrived on Earth.

### Source ejecta craters for the different SNC martian meteorites

Martian meteorites are impact ejecta by the bombardment of fragments of asteroids or comets onto the surface of Mars, and the three main types of martian meteorites are shergottites, nakhlites, and chassignites (SNC). If one can identify the source impact crater of martian meteorites, we could more accurately describe its geological evolution, and more effectively plan for future missions.

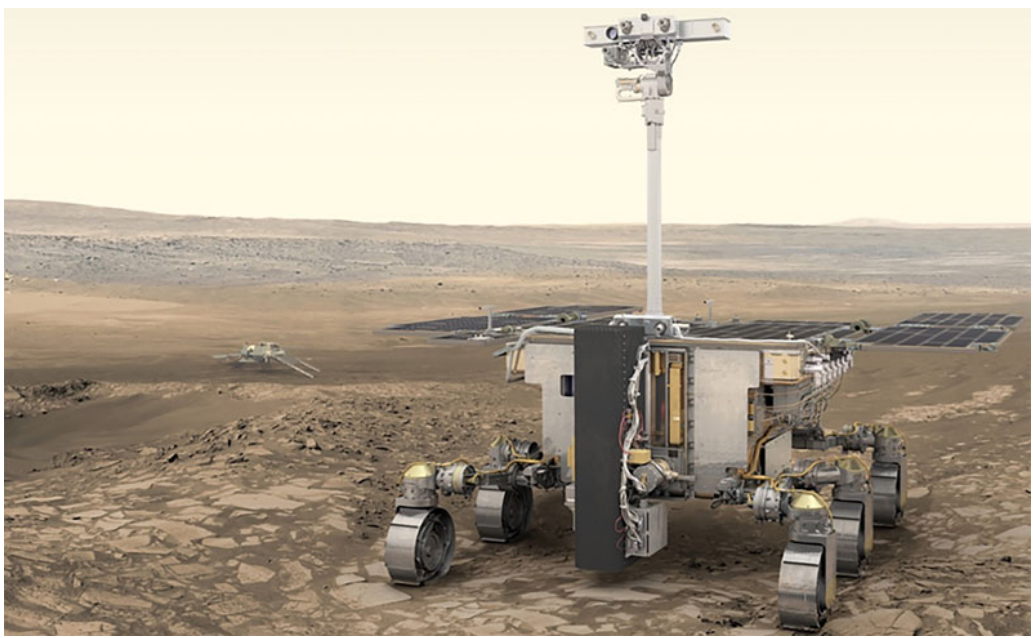

**Fig. 8.** Artist rendition of the Rosalyn Franklin ExoMars Rover.

Finding the source craters for our more than 140 martian meteorites is however very challenging. When we classify a martian meteorite on the basis of its mineral abundances and compositions, or trace element abundances, we are at a level of detail that is very hard to replicate from remote spectroscopy. It may happen for some meteorite samples, but we will learn about Mars from the meteorites as fragments of the crust, but not from known exact locations. Even if we are not sure exactly which craters they were derived from, they still give essential pieces of information about the planet's differentiation and the action of water in the crust.

### Noble gas composition of the martian atmosphere

Signatures of the 'ancient martian atmosphere' (based on, e.g. the martian meteorite Alan Hills (ALH) 84001) and 'modern martian atmosphere' (from, e.g. Northwest Africa (NWA) 7034) can be traced by noble gases. Trapped noble gases (Ar and Xe) in martian meteorites such as shock-produced glasses are in good agreement with MSL measurements on Mars ($^{40}Ar/^{36}Ar = 1900 \pm 300$, $^{36}Ar/^{38}Ar = 4.2 \pm 0.1$, $^{129}Xe/^{132}Xe = 2.5221 \pm 0.0063$) (Avice et al., 2018). Based on meteorite studies, $^{40}Ar/^{36}Ar$ on Mars is much higher and $^{36}Ar/^{38}Ar$ is lower than Earth's atmosphere, indicating enrichment in radiogenic $^{40}Ar$ compared to $^{36}Ar$, possibly due to solar wind-induced sputtering (Atreya et al., 2013). The $^{129}Xe/^{132}Xe$ ratio in martian meteorites and on Mars is elevated due to radiogenic $^{129}Xe$. This requires either very early loss from the atmosphere or storage in some reservoir until after atmospheric loss (Swindle, 2002).

### Noble gases from the martian surface and interior

Mudstone on the floor of Gale Crater analysed by MSL determined K-Ar age of $4.21 \pm 0.35$ billion years, representing a mixture of detrital and authigenic components in ancient rocks making up the crater rim (Farley et al., 2014). Furthermore, the surface exposure age determined by the cosmogenic isotopes $^{3}He$, $^{21}Ne$ and $^{36}Ar$ is $78 \pm 30$ million years, indicating surface exposure rather than primary erosion and transport. This absolute age data enables calibration of relative surface ages for crater chronology.

In addition to the atmospheric components, martian meteorites can also contain trapped martian interior components, which are more challenging to measure with the current Mars landers and rovers. Information we have on martian interior gases are primarily mostly from the meteorite Chassigny. Fluid inclusions in olivine in Chassigny have higher Kr and Xe content, and minimal radiogenic $^{129}Xe$, unlike the martian atmosphere (Swindle, 2002). Chassigny has also a much lower Kr/Xe than Mars' current atmosphere and a Xe isotopic composition close to that of the solar wind, suggesting an undegassed interior martian reservoir (Swindle, 2002). In addition to solar Xe, another interior component has been suggested due to the addition of $^{244}Pu$ fission component from the martian interior (Mathew and Marti, 2001). In summary, noble gases can be highly diagnostic when unravelling planetary formation and evolution.

### The martian surface and interior revealed by martian meteorites

The martian breccias, NWA 7034 and its many pairs, contain the most evolved martian igneous material and a high amount of volatiles including water. They are also the eldest martian samples yet identified. Water released by step-heating of the martian breccia showed concentrations of ∼3000 ppm, an order of magnitude above that of any other martian meteorite (Agee et al., 2013). Its composition is also enriched in alkalis and other volatiles relative to the SNC meteorites. The dating of zircons included within the breccia show ages around 4.4 and 1.7 Ga, suggesting an ancient initial crystallization age followed by a later major disturbance (Humayun et al., 2013; Bouvier et al., 2018). The most-precise analyses of zircons and geochemical modelling predict the existence of an evolved andesitic crust on Mars with a minimum age of 4547 Ma (Bouvier *et al.*, 2018). This would have had to pre-date the currently basaltic crust on Mars and was likely later destroyed by impacts and resurfacing.

### The shergottites

The shergottites make up the highest fraction of martian meteorites. In the current collection of martian meteorites, ~80% are shergottites, including basaltic, olivine-phyric and lherzolitic shergottites (martian Meteorite Compendium) (Wang and Hu 2020), as well as the recently identified augite-rich shergottites, NWA 7635 (Lapen *et al.*, 2017) and NWA 8159 (Herd *et al.*, 2017). Most of the shergottites have crystallization ages ranging from ~600 to ~150 Ma by the analyses of mineral isochrons of Rb-Sr and Sm-Nd (Nyquist *et al.*, 2001), and *in-situ* U-Pb dating of baddeleyite and phosphates (Niihara, 2011; Moser *et al.*, 2013; Zhou *et al.*, 2013). Two new augite-rich shergottites were however dated at ~2.4 Ga (Herd *et al.*, 2017; Lapen *et al.*, 2017). The relatively young ages of most shergottites (Váci and Agee, 2020) indicate that they were ejected from the Amazonian Epoch of Mars. On the other hand, the ejection ages of shergottites vary from ~1 to ~5 Ma except for the shergottite, Dhofar 019 at ~20 Ma (Fritz *et al.*, 2005). These variations suggest that there may have been several impact events on Mars that eventually delivered material to Earth.

Some work has been carried out attempting to constrain the source craters of the shergottites (Mouginismark *et al.*, 1992; Treiman, 1995; McFadden and Cline, 2005; Tornabene *et al.*, 2006) and other martian meteorites (Kereszturi and Chatzitheodoridis, 2016). Some candidate craters have been proposed – examples being the Mojave Crater, forming <5 Ma ago, with a width of ~55 km, on the 4.3 Ga old terrain of Mars. It was proposed as the ejection source for the shergottites due to its very similar reflectance spectra to the shergottite Queen Alexandria (QUE) 94201 and Shergotty in the wavelength range of 1–2.4 cm (Werner *et al.*, 2014). However, the size of Mojave Crater (Fig. 9) is one order of magnitude higher than that simulated by the launch dynamics (Head *et al.*, 2002). The very young age of Mojave Crater (Fig. 7) also requires better constraints because of the low accuracy of crater chronology for the young craters without ground-based calibration (Hartmann and Neukum, 2001). Moreover, the host terrain under the Mojave Crater is substantially older than most of the young shergottites. Future martian sample return missions will contribute greatly in clarifying this issue (Beaty *et al.*, 2019).

### Alteration in the martian meteorites

Alteration minerals and textures have been identified in martian meteorites, namely in the nakhlites. The nakhlites are clinopyroxenites (Treiman, 2005). Alteration mostly occurs along olivine fractures and in mesostases. A martian origin of alteration in the other types of martian meteorites are controversial, although carbonate globules in ALH 84001 are generally accepted as extraterrestrial (Treiman, 2021). Carbonate was first suggested to occur in a martian meteorite by (Carr *et al.*, 1985) using C isotopic measurements during stepped combustion of Elephant Moraine (EETA) 79001 martian meteorite. Carbonate was later discovered in Nakhla as Mn-rich siderite (Chatzitheodoridis and Turner, 1990) in an assemblage with Ca-sulphate and halite. Additional evidence came from Gooding *et al.* (1991). Since then, the discoveries of alteration phases have been rapidly increasing in number. The list of secondary minerals includes siderite ($FeCO_3$), anhydrite ($CaSO_4$), NaCl, ferric saponite clay ($Ca_{0.25}(Mg,Fe)_3((Si,Al)_4O_{10})(OH)_2 \cdot nH_2O$), hematite ($Fe_2O_3$), oxyhydroxides (ferrihydrite $Fe_{10}^{3+}O_{14}(OH)_2$), serpentine (($Mg,Fe)_3Si_2O_5(OH)_4$), and opal ($SiO_2 \cdot nH_2O$). Their discoveries are important for understanding martian weathering processes which have implications for interpreting the paleoenvironment of Mars. Moreover, they may hint at potential ecological niches in the subsurface of Mars, such as vesicles formed in mesostasis formed by bolide impact processes (Chatzitheodoridis *et al.*, 2014) or for instance networks of microcavities of magmatic clays serving as microreactors for prebiotic chemistry (Viennet *et al.*, 2021). Similar types of alteration on Mars provide direct comparisons between Earth-based samples and Mars prior to the eventual return of samples from the planet.

### Martian Moon eXploration sample return

The Japanese Aerospace Exploration Agency has a Mars moon sample return mission planned for launch in 2024 to Phobos and Deimos; the Martian Moon eXploration Mission (MMX). The orbits of these satellites suggest that they formed either from large impact ejecta from Mars that then accumulated, or at the same time as proto-Mars itself (Bagheri *et al.*, 2021), although they share morphological and spectral characteristics with asteroids (See *Early Mars* section). Subsequent to their formation, the martian moons accreted materials transported from Mars, probably delivered from cataclysmic impact events on the planet.

Phobos and Deimos may have therefore been contaminated by microorganisms if there was life in the martian subsurface. The mission will collect samples from Phobos, with successive flybys of Deimos. If putative martian life is transferred to Phobos or Deimos, what unique challenges will arise for a sample return mission? According to previous assessments of microbial contamination risk for samples collected on the martian moons (Fujita *et al.*, 2019; Kurosawa *et al.*, 2019), the probability that a single unsterilized particle occurring within samples can be maintained within the internationally agreed acceptable limit of $10^{-6}$, provided that the mass of samples is below 100 g, and that they are taken within a depth of several tens of centimetres from the surface. The extant microbial density on the martian moons (if in fact Mars was inhabited) is subject to the rates at which microbes are inactivated by high-speed impacts of martian ejecta on the martian moons and by cosmic radiation after accretion onto the surface. The above assessment has been conducted based on laboratory simulation experiments. However, there still remains uncertainty with these results. For this reason, it is strongly recommended by the Committee on Space Research (COSPAR) that a re-evaluation is made before the return of martian moon samples to Earth. For this purpose, we have started new follow-up experiments to constrain these uncertainties.

### Sampling asteroids versus Phobos

The Japanese Aerospace Exploration Agency also returned the samples from asteroid 162173 Ryugu in December 2020 with the Hayabusa-2 spacecraft, the first successful sample return mission to a carbonaceous asteroid. There are several challenges with a sample return mission from the martian moons that differ to the sample return from asteroids. From a technical viewpoint, it is a great challenge implementing a sampling method that complies with the above criteria in the absence of detailed information on the state of the regolith on the martian moons. Although gravity on the martian moons is much smaller than that of Mars or Earth's moon, it is several orders of magnitude larger than that of asteroids (Yang *et al.*, 2019). This requires a more-tailored propulsion system directed towards the martian moons during touchdown and lift-off. The exhausts from the propulsion systems may disturb regolith on the martian moons, and potentially accrete regolith onto the spacecraft by electrostatic attraction. Regolith scattering can also occur by touchdown of the landing gears and sampling. Adhesion of regolith to the spacecraft will substantially affect the bulk amount of the martian moons samples brought back to Earth, potentially resulting in a higher probability of microbial contamination. In order to avoid such a risk, it is not only necessary to suppress the regolith scattering by optimizing the layout and operation of the propulsion system, by minimizing the touchdown velocity, and by devising the sampling mechanism so that it can quietly collect samples (without creating turbulence), but also to identify the area of regolith adhering to the spacecraft. The spacecraft system and its operation plan for the MMX mission are currently designed according to the above policies with the aid of the analysis on the regolith touchdown, sampling and lift-off dynamics.

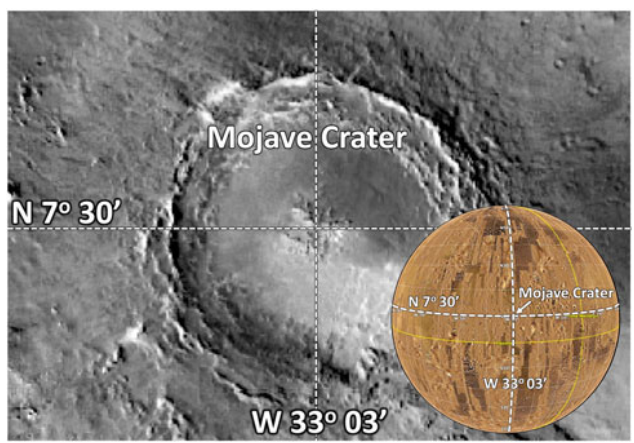

**Fig. 9.** Mojave crater (58 km across) on the Xanthe Terra plain of Mars. NASA/Caltech/Arizona State University..

### Mars 2020 sample caching

*Perseverance* successfully collected and cached its first rock sample from the martian surface on 1 September 2021. This was the first sample collected for return from the surface of another planet. There are 43 sample containers in a carousel for sample caching in titanium hermetically sealed tubes. Five unsampled witness tubes are included to identify any possible cross-contamination from Earth by the sample caching process. Samples will be deposited locally on the martian surface of the planet to be picked up by a fetch rover or perhaps even delivered by *Perseverance* in a follow-up mission towards the end of the decade.

The first sample collected by *Perseverance* belongs to a rock named 'Rochette'. The sample caching system contains a drill used for abrasion of the surface of rocks to reveal their subsurface, followed by actual core drilling and sampling. The acquired sample is a drill tube of rock that weighs 15 g (core diameter of 1 cm and length of 5 cm). Primary to the selection of samples, analytical information can be acquired with the instruments such as SHERLOC and PIXL. The selection of the samples will be based entirely on the remote detection of areas of interest; once an interesting sample is located, drilling and cashing operations will be performed.

### Mars human exploration

#### What are the essential requirements for humans to live and thrive on Mars?

Eventually, sooner or later, human missions to Mars will occur. If they are to constitute footholds rather fleeting jaunts, martian resources must be utilized to support humans in a lifestyle that is both sustainable and, just as importantly, comfortable. It is clear that sustainable settlement in a martian environment must be systematic and as complete as possible to maximize the use of *in-situ* resources, and minimize reliance on Earth through a long, expensive, and slow-reacting supply chain. Resource technologies designed to enable the autonomy of humans on Mars (so that they do not rely on Earth), as well as to characterise the martian environment, are under development (atmospheric composition, hydrology, mineralogy, and geological history). A comprehensive review of martian *in-situ* resource utilization is given in Ellery and Muscatello (2017) and details on sustainable *in-situ* resource utilization in Ellery (2020). Agriculture, for the generation of food, and the recycling of nutrients to substitute for biogeochemical cycles on Earth is essential – of course, martian soil may be utilized or hydroponics and aeroponics provide the means to render agriculture independent of soil – but in all cases, nutrient recycling will be required.

Key technological challenges arising from human space travel to the exploration of Mars include (i) survival during our trip towards Mars in the frame of investigations on the stability and level of degradation of space-exposed biosignatures such as pigments, secondary metabolites and cell surfaces in contact with a terrestrial and Mars analogue mineral environment; (ii) space radiation, because it has been demonstrated that the particle radiation environment on Mars can vary according to the epoch concerned and the landing site selected; (iii) human-safe landing and return systems; (iv) life-support and learning to live sustainably on Mars (in synergy with technologies developed for extreme environments on the Earth and the Moon). However, we still have to address a number of important questions such as what are the limits to interplanetary human travel to Mars, which technologies have still to be developed, and what studies and human simulations should be performed? Substantial research and development is required to provide the basic information for appropriate integrated risk management, including efficient countermeasures and tailored life support. Methodological approaches could include research on the International Space Station (ISS), on robotic precursor missions to Mars, in ground-based simulation facilities, as well as in analogue natural environments on Earth (Foing *et al.*, 2011*a*, 2011*b*).

#### Resource utilization on Mars

One key resource is the carbon dioxide atmosphere as it has long been proposed that carbon dioxide can be converted into methane using hydrogen feedstock via the Sabatier reaction (Zubrin *et al.*, 2013). Methane may be used as a propellant, which combined with an oxidizer, can provide the basis for refuelling Mars Ascent Vehicles and rover sorties during in-situ exploration activities. Water ice on Mars can provide another resource for life support, hydrogen feedstock, and oxygen for respiration. Acquiring water ice, however, will require extraction from the martian subsurface through drilling/mining techniques. Finally, raw regolith may be exploited by combining with binders (such as plastic or salts) for the construction of habitation shells using 3D printing (Cesaretti et al., 2014). There is much we can leverage from Mars in constructing the internal environment for a martian habitat (Ellery, 2020).

### Technologies for resource utilization

To make Mars our home, an industrial capacity from martian materials should be built. Machines will provide the means of production to construct this industrial capacity. We need to build excavator rovers, comminution and beneficiation machines, unit chemical processors, milling machines, 3D printers, assembly manipulators, and other machines that convert raw material into useful products. These machines are all kinematic machines – the fundamental components of the kinematic machine are the electric motor, its control system and its power supply (Ellery, 2016). These technologies will require a basic set of materials including iron (from hematite), nickel, cobalt, tungsten and selenium (from nickel-iron meteorites), silica and silicon (from silicate minerals), plastics (from a mixture of $H_2$ and CO) and acid reagents (from salts). The core unit chemical processor is the Metalysis FFC (Fray-Farthing-Chen) process supported by a range of acid-based and carbonyl-based peripheral processes (Ellery, 2017). Combined with 3D printing technology (Ellery and Muscatello, 2017), this provides a powerful suite of techniques for universal construction capabilities including the construction of electric motors and computational electronics (Ellery, 2016), the latter based on neural network architectures (Prasad and Ellery, 2020). Human exploration can leverage these same components to support energy generation through solar concentrator-thermionic conversion and energy storage through motorized flywheels. We can thus construct an industrial infrastructure of machines as the fundamental requirement for self-sufficient habitability.

### Bio-materials for in-situ resource utilization

Synthetic biology and applied microbiology can be used to produce resources in space (Rothschild, 2016). For example, carbonate biomineral-producing microbes are used on Earth in civil engineering, art restoration, soil improvement, bioremediation, and $CO_2$ sequestration (Reddy, 2013; Anbu et al., 2016), as well as space. Calcite production is common among microorganisms and can be linked with different types of taxa and metabolisms (Boquet et al., 1973). Bacteria, such as *Streptomyces* species, promote the precipitation of calcium carbonate by activating specific metabolic pathways. For biotechnological applications, the genera *Bacillus* and *Sporosarcina* are the most studied (Han et al., 2019), but knowledge gaps remain about the microbes best-suited for each type of application, and how environmental conditions affect the characteristics and yields of biominerals. Terrestrial bacteria found in calcium-rich natural environments on Earth could be used on the Moon or Mars for the production of biobricks for constructing buildings (Santomartino et al., 2020). Some strains that precipitate calcium carbonate or gypsum (depending on the substrate provided; Cirigliano et al., 2018) are also able to produce melanin that protects against UV and ionizing radiation.

Biomineralization processes have been proposed as a means of cutting the construction costs on the Moon and Mars (Cockell, 2010; Dikshit et al., 2020; Kumar et al., 2020). More work is needed on the feasibility of the use of Mars dust and regolith for such a proposed in-situ resource utilization, e.g. testing this process under Mars-like conditions. One limitation may be the various salts that are known to inhibit or prevent microbial metabolism, including perchlorates and chaotropic salts, sulphate salts, and those that create multiple stresses and polyextreme conditions (Fox-Powell et al., 2016; Hallsworth, 2019; Benison et al., 2021). Work is therefore needed to identify strains tolerant to salt-induced stresses and optimize culture conditions to mitigate against the underlying cellular stress mechanisms.

### Human landing site on Mars

At the current time, there is a major community-wide discussion taking place regarding the required/desired attributes of the future human landing site, and possible locations on Mars that have those attributes. At the present stage of the discussion, there are multiple sites actively under consideration. Some key factors include elevation (which has to be below a certain level), minimize landing site hazards such as slopes and large rocks, attributes related to the purpose of mission (such as science), access to water ice, and planetary protection. Some candidate regions for SpaceX's *Starship* are Phlegra Montes, Erbeus Montes and Arcadia Planitia (Golombek et al., 2021).

### Terrestrial simulations of human missions to Mars

There is a number of ground-based simulation facilities in analogue natural environments on Earth. Building on the EuroGeoMars 2009 campaign (Foing et al., 2011a, 2011b), the ILEWG EuroMoonMars programme since 2009 has included research activities for data analysis, instrument tests and development, field tests in the MoonMars analogue, pilot projects, training and hands-on workshops, and outreach activities. Field tests have been conducted in ESTEC (European Space Research and Technology Centre) and EAC (the European Astronaut Centre) of ESA, as well as at the Utah's MDRS station (Mars Dessert Research Station), Eifel, Rio Tinto (Spain), Iceland, La Reunion (an island in the Indian Ocean that belongs to the French Republic), LunAres base at Pila Poland and the HISEAS base in Hawaii.

### Simulating human experience on Mars using virtual reality

Humans can experience Mars using virtual reality. Current virtual reality engines can elevate the quality and immersion of Mars with the high-fidelity mechanics, interactions and continually evolving design principles.

By using Mars topological and in-situ data from current orbiters and rovers, respectively, extrapolation of a 3D virtual reality martian landscape can be made. All touch points within the experience (locations, mineral outcrops, weather, caves, tools, etc) can be scientifically accurate on a 1:1 scale, and provide transferable knowledge beyond the virtual reality experience (Fig. 10).

In virtual reality, humans can navigate the hazards of the martian surface, and use simulated rover playloads, spacesuits and handheld scientific instruments to conduct experiments on the surface of Mars and discover the viability of interplanetary human expansion.

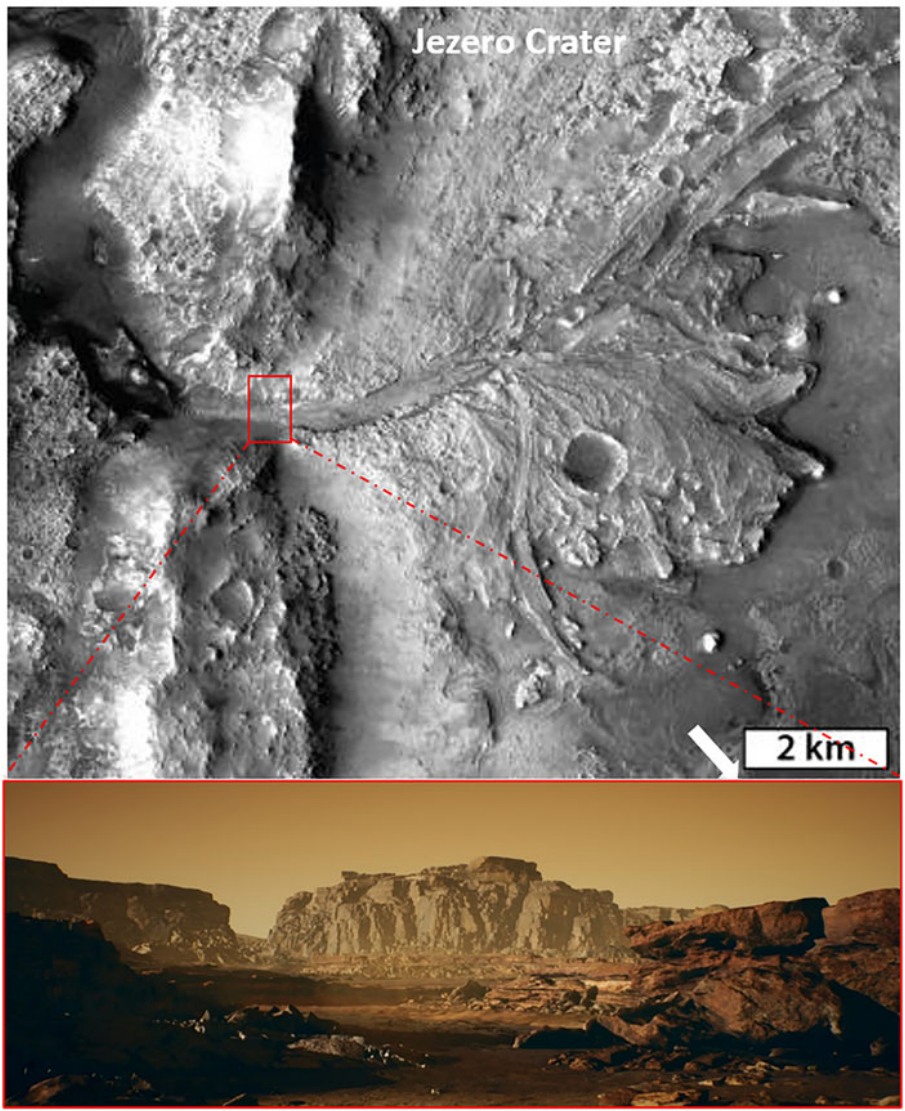

**Fig. 10.** In-engine virtual reality (VR) terrain of the Jezero Crater looking into the river delta (bottom) extrapolated from Mars Reconnaissance Orbiter imagery (top). Top image courtesy of NASA/JPL/JHUAPL/MSSS/ Brown University. With imagery from the *Perseverance* rover, the VR engine can be refined for more accuracy. Arrowed is the current approximate location of Perseverance.

### Mars human settlement is 'ante portas': are we ready for it?

Human settlement on Mars is the next coming human milestone. However, societal risks emerge on a global scale (Shaghaghi and Antonakopoulos, 2012). Quantitatively estimating how to cope with those radical societal changes primarily invents a way of measuring and monitoring the societal readiness towards Mars human presence, from Earth's and Mars' points of view, and then identifying and remediating the risks that will arise for humanity. Such a tool could be quantitatively expressed with a risk index, such as the Mars Risk Index (MRI), a data-driven index estimated under the MRI Framework which monitors the societal readiness and identifies the risks that arise for humanity. It assists all stakeholders to orientate themselves as Mars human settlement takes shape. It is envisaged that two main sub-indices will allow the evaluation of the MRI:

(a) The Mars Earth Risk Index ($\mu$RI:e) that measures Earth's societal readiness shifting towards the Mars human settlement.
(b) The Mars Ares Risk Index ($\mu$RI:a) that measures Mars' future societies structures readiness to cope with Earth's existing societal structures.

To explain briefly the MRI's methodology (Fowler *et al.*, 2004), the analysis of societal and political risks will take firstly place, based on the relationship between frequency and the number of people suffering from a specified level of harm in a given population from the realization of specified threats. The quantitative risk assessment approach is based on the estimations of the pair of frequency (f) and potential consequences, and the typical number of fatalities (N), resulting in a cumulative frequency (F) of events that has N or more fatalities, visually compared through sets of curves, commonly referred to as F-N curves. With the use of such curves, judgements will be made. To further establish this framework, possible threats, the available assets and the estimated impacts must be taken into account.

*Minimizing human risks for settlements on Mars*

Human settlement of Mars will be a consorted international endeavour. Its stakeholders will span all societal pillars such as, nations, countries and their governments, international organizations, religious groups, and of course, people (i.e. individuals).

As a consequence of the societal risks that emerge on the global scale, unstable environments initiating domino effects might be created that will shake our societies' structures. It is our responsibility to adapt our societies to this endeavour, which is by far larger than, i.e. the discovery of the New World. MRI can help all stakeholders orientate themselves as Mars human settlement and colonization takes shape, alert them for changes that could mitigate those risks and confront the challenges.

**Planetary protection**

Mars has been categorized as a planetary protection restricted class Category V, relating to the return of space craft or space components. This means that where scientific opinion is unsure, destructive impact upon return to Earth is prohibited. Containment of all returned hardware which directly contacted the target body, and any unsterilized sample returned to Earth, is also required. The same is also required for the forward contamination of Mars itself.

*If either extinct or extant life is discovered on Mars, should the UN resolutions on planetary protection be amended?*

If any indications of Mars life are found in early assessments, one has to assume that life continues to exist on Mars until more data are collected.

Currently, the international policy on planetary protection as maintained by COSPAR states explicitly that any decisions about novel mission activities should be based on scientific advice from the COSPAR Member National Scientific Institutions and International Scientific Unions. This includes a reassessment on the status of Mars if evidence for indigenous life is discovered. The current COSPAR guidelines for analysis of samples returned from Mars focus on the potential for false-positive results, because a false positive might result in unnecessarily increased restrictions on Mars exploration.

However, any evidence for past Mars life would significantly increase concerns about the potential for false-negative detections of extant Mars life. If samples from Mars provide evidence of indigenous life in the distant past, it is not valid to assume all Mars life subsequently became extinct. Due to the potential for hazardous interactions between Mars life and Earth life, the detection of past life on Mars would appropriately lead to increased restrictions on the exploration of Mars and return of Mars' material to Earth. The 2020 pandemic of COVID-19 illustrates some potential consequences resulting from false-negative detections of hazardous organisms. Planetary protection was established and needs to continue in effective implementation, to prevent space exploration causing this sort of pandemic scenario. A thorough assessment of various modes of inactivation and containment of martian sampling is ongoing (Craven *et al.*, 2021).

*Regarding the contamination of Mars itself, should the planetary protection protocol between the Moon and Mars be any different?*

The current COSPAR guidelines for missions to Mars are significantly different from those for the Moon, because Mars is considered to provide potentially habitable environments for indigenous Mars life. In addition, some environments on Mars, particularly in the subsurface, could be habitable for Earth organisms. Therefore, stringent limits are placed on the number of Earth organisms permitted on outbound missions to Mars, to reduce the potential for Earth organisms to be delivered to, and persist, on Mars.

There is no concern for contamination of the lunar surface, because the environment of the Moon is known to be uninhabitable by carbon-based life that functions in physical environments similar to the surface of Earth. In addition, for the purposes of planetary protection, Earth and the Moon are considered to be one system, which ensures no travel restrictions between these two bodies will be emplaced.

*Planetary protection is inextricably linked with human health*

The microbiomes of modern humans living in urban settings are less diverse than those of ancient indigenous peoples as shown by a range of recent seminal studies (De Filippo *et al.*, 2010; Clemente *et al.*, 2015; Eisenhofer *et al.*, 2020; Sprockett *et al.*, 2020; Weyrich, 2021). Diversity within the human microbiome is reduced when our lifestyle isolates us from contact with biodiverse soils (Blum *et al.*, 2019) and the microbiomes of astronauts are known to become impoverished due to isolation and from Earth's biosphere (Sugita and Cho, 2015). Therefore, we can expect that humans living on Mars will not enjoy optimal microbial biodiversity. The additional use of ethanol-containing (or other) wipes to cleanse the skin further reduces the biodiversity of skin. These factors, along with the use of antibiotics or other drugs that perturb and diminish the microbiome, will reduce the usual competition within the microbiome. Biodiverse, healthy and climax microbial communities are most likely to suppress the proliferation of pathogens or pathogenic events (Cray *et al.*, 2013).

Conversely, damage to microbial communities on and within the human body that are perturbed and diminished results in the creation open habitats (places that are resource-rich and promote proliferation of and competition between microbes) (Cray *et al.*, 2013). These include excessively cleaned skin, wounds and surfaces of the gastrointestinal tract after use of antibiotics. Many of the potentially pathogenic microbes that occur with the human microbiome have a robust stress biology and are therefore selected for by the use of skin disinfectants (Suchomel *et al.*, 2019). They are also more likely to be capable of survival if faecal waste is used as a fertilizer and so can re-enter the body via the oral-faecal route due to consumption of contaminated foods (see below). In the event that any human(s) become(s) infected with a highly contagious microbial pathogen, it has yet to be determined how to prevent an epidemic within the human population of Mars. In the unfortunate event of a human fatality on Mars, there is likely to be a contingency plan to recover the body to Earth. However, the time delay until this occurs raises real questions about planetary protection and safeguarding the health of the other people remaining in the habitation(s).

Indeed, there are ethical, technical and scientific dilemmas about humans and microbes living on Mars that have yet to be resolved. In relation to microbes on Mars, it will be difficult to live with them, yet we cannot live without them.

### Microbes for human medicine on Mars

Human Exploration needs to address potential issues related to the health and illness of astronauts. Therefore, a broad range of basic medicines is needed within the payload, as well as the technology to produce them on demand whilst in transit or in the habitation(s) on Mars. This includes any unexpected medicines needed (different from those brought from Earth) and supplementing any exhausted medical supplies. The miniaturization of portable scientific instruments for use during space exploration has also benefited biotechnology on the Earth (especially the development of micro-bioreactors, mini-microscopes, mini-DNA/-RNA sequencer). Many active pharmaceutical compounds are of plant origin but it is expected that if plants grow on Mars, they will be intended for food. Synthetic biology, via the use of metabolically engineered bacteria and yeasts, can be used to produce (some of these are already produced commercially on Earth) complex molecules such as antibiotics and the antimalarial drug (Ro *et al.*, 2006; Perez-Pinera *et al.*, 2016; Liu *et al.*, 2018; Malico *et al.*, 2020). A culture collection of strains (kept freeze dried in small vials) to use for drug synthesis provides a choice for humans on Mars to be able to select the appropriate microorganism for growth in a micro-bioreactor and produce the required medicine. Currently, 3D printing technology can produce personalized medicine containing one or more active ingredients; this is already a reality on the Earth (Reddy *et al.*, 2020).

Mars-related diseases, unknown on Earth, could arise and so plasticity of medical solutions and synthetic biology capability are required because there are considerable unknowns and astronauts need to have the capacity to respond and adapt to any eventuality. However, microorganisms in nature are metabolically versatile and are, for example, able to degrade chemically diverse hydrocarbons and other substrates (Timmis, 2002; Cray *et al.*, 2013) and produce an indefinite number of secondary metabolites (Cray *et al.*, 2015; Kai, 2020). This capacity has value in metabolically engineering strains to produce biofuels and plastics (Nielsen *et al.*, 2013; Ko *et al.*, 2020) as well as pharmaceuticals (see above). Another example is provided by *Saccharomyces cerevisiae* that can produce the precursors of analgesics such as morphine (Pyne *et al.*, 2020) and cannabinoids (Luo *et al.*, 2019). Unresolved issues of using microbes on Mars for this purpose are whether mutations might occur, and whether drugs can be produced in sufficient quantities and whether they can be readily purified. Apart from its use in drug production, yeasts might be useful to produce vanilla and raspberry flavours for use as food additives to help avert a sense of homesickness and to add to the *joie de vivre*.

## Mars and society

### Society, ethics and theology on Mars exploration

In regard to societal, ethical, and religious aspects of Mars exploration it is useful to distinguish between *human* exploration and *scientific* exploration. This is often described as a science versus exploration scenario in space exploration (Metzger, 2016). Human Mars exploration can be understood as an expansion of Earth society, while scientific exploration enables Earth society to learn about its place in the Universe. This perspective has been relevant for philosophers and theologians for centuries. One example of such debate is the question of life on other planets. Thomas Payne as well as John Calvin have discussed the theological aspects of life on other planets (Weidemann, 2014). For Immanuel Kant, life on other worlds was certain and this 'knowledge' informed his philosophical stance.

While scientific exploration of Mars via robotic probes and landers has been going on since the beginning of human space exploration, a human mission to Mars is still far-reaching. It is interesting to note, that the possible presence of life on Mars is one of the main drivers as well as one of the main obstacles in Mars exploration. The debate on planetary protection of Mars and its impact on mission profiles boils down to the ethical question of how we as humans should relate to possible extraterrestrial life. This ethics question has a wide range of answers: from very human-centred to biocentric perspectives that give all life the same value, and even, perspectives that ascribe non-inhabited areas of space the same value as inhabited areas. A religious perspective on this debate boils down to the question 'what role humans play within God's plan or creation'. If we are at the centre of it, we still might be obligated to care for the rest of the species and not just profit from them. For Mars exploration, this could mean an obligation to care for possible life on the Red Planet. Chris McKay and Robert Zubrin have discussed this in relation to the settlement of Mars (Waltemathe and Hemminger, 2019). While McKay argues from a biocentric stance to change the environment of Mars in such a way as to enable indigenous bacterial life to prosper and enable humans to learn about life itself, Zubrin basically argues to freeze indigenous life and keep it in the lab while making the planet liveable for humanity. He employs a strong anthropocentric view of planetary colonization. Brian Green of the University of Santa Clara, in an opinion piece for CNN, questioned both positions from a philosophical as well as a theological point of view, and restructured their discussion to find a middle ground. All these discussions aim at giving the current planetary protection ruling an ethical framework. The current regulations are meant to protect scientific evidence from contamination through exploration. Their rationale is protection of science. One could argue, however, that this rationale needs an ethical and societal framework that includes more than science as a rationale (Sherwood *et al.*, 2019).

The theological implications of life on Mars and the societal response to the discovery have received scholarly attention (e.g. Vakoch, 2013). The broader societal benefits of the planned robotic and potentially astronaut missions to Mars however relate to its immense inspirational and educational values and potentials. Such efforts can attract society and in particular young generations to expand the limits of our knowledge, to evoke interest in the study of sciences and related disciplines, draw attention to the values of international, peaceful cooperation in space, and even to the possibility of creating a sustainable human presence in the solar system beyond Earth. Using Mars as a teaching context, engaging students and general public with exploration efforts using the ever-popular Red Planet are all possible directions for science dissemination and popularization, engagement and even for correcting the misconceptions about science and space exploration (Capova *et al.*, 2018).

### What is the motivation for space exploration?

Human exploration of Mars often also has another motivational aspect: survival. Freeman Dyson frames it for interstellar exploration as assurance against even the worst imaginable of natural or man-made catastrophes that may overwhelm mankind within our Solar System (Dyson, 1968). Dyson also argues that a space colony would achieve 'total independence from any possible interference by the home government'.

William E. Burrows in 'The Survival Imperative' (Burrows, 2004) also argues for safeguarding humanity not only through space exploration but also in space. He draws on the ideas of O'Neil and Sagan and points out the pros and cons of their concepts. This can be directly applied to possible Mars settlements. With SpaceX planning such settlements for the near future, this perspective becomes immediate.

If one is to understand the motivation for Mars exploration for safeguarding humanity's future in nearby and distant solar systems to save us from a cataclysmic event that might destroy our habitat on Earth, the religious connotation of Noah's Ark immediately comes to mind. This story that is part of the Jewish, Christian and Muslim tradition (e.g. Gen 6–11 in the biblical text, but equally present in the Quran in different suras) is part of the religious and cultural foundation of European, Mid-Eastern and American societies. Flood mythology, however, can be found in all major religious traditions. As a consequence, Mars exploration scenarios would be well advised to take these religious and cultural connotations into account and relate to their specific challenges and opportunities. Another motivation is establishing new resource bases for humanity, thereby lightening the load on Earth's resources. Robert Zubrin in 'Entering Space' builds his argument in a similar way (Harris, 2000). Exploration as part of the human condition is something that is also argued in regard to space exploration and therefore to Mars exploration. The concept of 'Manifest Destiny', for example, is being invoked by the proponents of space exploration. William E. Burrows writes 'At the heart of it all, as usual, (were) the core of dreamers…who steadfastly believed it was their race's manifest destiny to leave Earth for both adventure and survival' (Burrows, 2010).

### What has shaped public opinion about Mars?

Public attitudes about Mars are shaped by narrations created on Earth by science fiction authors, scientists, philosophers and politicians (Messeri, 2016). Historically, the Red Planet figures the public imagination about other worlds and its inhabitants. A prominent example is the 1897 novel 'The War of the Worlds' by H.G. Wells that has shaped the envisioning of an alien invasion and dominated a century-long narrative history. Today's explorers and science communicators are presented with a challenge of how to disseminate good science, its concepts and results in parallel to the many science fiction stories that are entertaining yet not always science-conformant. Here, for example, the robotic missions designed to find evidence of possible existence of life in habitable niche environments, e.g. *Perseverance* and *Rosalin Franklin*, have the capability to capture popular imagination as they seek answers to the questions: was there life on Mars, and what was it like?

### Societal benefits of Mars exploration

Human and robotic space exploration responds to the deeply rooted quest of humankind for answering questions on the origins and nature of life in the Universe and extending human frontiers. More down-to-earth considerations are however important driving forces behind the actual decisions for investing in exploration programmes.

In the 1950s, geo-political considerations were the main driver for space exploration. What eventually evolved into a cold war space race to the Moon did and still continues to accumulate massive economic and societal benefits far beyond and above the outcomes initially envisaged. A study commissioned by the European Union in 2015 estimates that 6% of Europe's entire economy (including non-space) is directly dependent on space infrastructure and accounting for around 14 million jobs. Satellites are able to address about 60% of the 57 essential climate variables (ECVs), with several exclusively derived from satellite measurements. ECVs are critically contributing to the characterization of Earth's climate, providing a picture of climate change at a global scale. The Climate Change Initiative of the ESA generates consistent, long-term and global data records for 21 key ECVs.

With the satellite applications fully integrated into society and becoming mainstream, exploration of outer space and Mars still entails significant short-term and long-term benefits for society. This is largely recognized by the International Space Exploration Coordination Group (ISECG), a coordination mechanism of space agencies from 25 countries. Largely building on ISECG's vision and roadmap with horizon goal human exploration of Mars, ESA's European Exploration Envelope Programme (E3P) is implementing a benefits-driven European Space Exploration Strategy, prioritising the scientific, economic, inspirational and international cooperation dimensions of exploration. Earth orbit, Moon, and Mars (ESA's three exploration destinations) should be seen as part of a single coherent programme expanding in a step-wise manner the places where humans will work and live in space.

### Earth sustainability

In addition to the discoveries, advancement of science, technological development, and economic return can space technologies address other societal issues and assist in achieving sustainable development on Earth? The United Nations' Office for Outer Space Affairs (UNOOSA) has recognized space as a driver for the sustainable development and acknowledged the unique potential of space technologies in leveraging innovative solutions and technological developments. While the example of the Earth observations and geospatial data have already earned their recognition in contributing to the UN Sustainable Development Goals (SDG, a universal call to action to end poverty, protect the planet and ensure that all people enjoy peace and prosperity by 2030), the actual potential of space applications in supporting the SDGs is much wider. One example is the technology developed for analysing the atmosphere of Mars that now helps to cut greenhouse emissions on Earth (ESA SDG Catalogue 2018). Space activities, among those of ESA and other space agencies, are an important tool serving development (Duvaux-Béchon, 2019). The challenge remains in utilising those technologies and maximising their potential to make life on Earth more productive, clean and sustainable, securing safe future for our planet and human generations to come.

### Knowledge transfer from space exploration

The ISS, continuously crewed since 2000, shows the benefits and potential of human activity in low Earth orbit, from basic science to innovation and knowledge fuelling emerging terrestrial applications in fields such as air filtration, protein crystal growth, robotic surgery, water recycling, materials, fluids and combustion (Thumm *et al.*, 2012). The stringent challenges of space exploration to the next destinations are an accelerator for innovation. Exploration missions require the development of cutting-edge technology in areas such as advanced robotics, artificial intelligence and additive manufacturing (3D printing), and to enable next-generation life support, habitation and waste management systems for human exploration. This provides opportunities for other sectors to partner with the space sector on joint research and development. At the same time, scientific discoveries in weightlessness enabled by human exploration are applied widely. The results are touching every aspect of everyday life, from health and medicine, public safety, consumer goods, to energy and the environment, industrial productivity and transportation.

The process of developing a mission such as ESA's ExoMars 2022 mission sending the *Rosalind Franklin* Rover to search for life in the martian subsurface is already sparking spill-overs to the terrestrial economy. For example, welding techniques developed for the mission are also used to manufacture aluminium cans, with potential to save 12% on raw materials. A report commissioned by the UK Space Agency estimates that Rexam plc, a British-based multinational consumer packaging company and leading manufacturer of beverage cans, could have saved £242 million on raw materials in 2014 had the company implemented the technique. Mars Sample Return ground- and flight elements that ESA is contributing to an international end-to-end sample return capability will eventually allow scientists to analyse samples in terrestrial laboratories. Mars Sample Return will require the development of ground segments (such as the receiving and curation facilities) and a qualified workforce to handle the samples. This has the potential to push the boundaries of European technological achievement, directly creating thousands of high-tech jobs and spin-off industries to the benefit of the wider economy. Through the inspirational and motivational value of aiming high and beyond our current grasp, Mars exploration activities will result in increased participation of young people in STEM education and careers, thereby increasing their contribution to Europe's knowledge economy and capacity to address global challenges in the future. For example, Mars climate was once similar to the Earth's but has gone through massive changes. Importantly, Mars climate research may help better understand the past history and predict future changes in our Earth's climate (Read *et al.*, 2015).

### Potential pitfalls of becoming an interplanetary species

If the arguments on the motivation of exploration accurately describe human nature in regard to exploration and discovery, then perhaps space exploration is the ultimate form of human destiny, opening up the borders of our own planet. This can be made compatible with major religious traditions including views on humanity's role in creation. It certainly would serve as a motivation, but also bring with it all the cultural problems that historically, exploration has resulted in on planet Earth. The question arises of how to mitigate the negative effects of such concepts when trying to apply them to space exploration. Currently, a shift in language can be observed, addressing the difficult concepts of the past and the expressions they have shaped. Space colonies are different from space settlements in the potential negative historical connotation of words such as *colonization*, *exploitation* or even *mining*. Crewed vehicles describe the social reality of contemporary space explorations better than manned missions of the past. Grounding exploratory motivation in the space between scientific curiosity and human experience may serve to mitigate the problems. Religious and philosophical traditions can serve to motivate exploration on one hand but also as a careful reminder of the problems that we were, and still are, connected to it. Gaining knowledge about the Universe maybe the more-desirable first step, which also fuels human exploratory drive but with caution derived from more information.

### Epilogue

Mars could have been more Earth-like before. The similarities between Noachian Mars (∼3.7–4.1 Ga ago) and early Earth have implications for the astrobiology of Mars, not least that traces of extinct martian life might remain at the planet's surface or subsurface. The tenuous atmosphere of present-day Mars evolved from the depletion of a thicker atmosphere. This drove a water cycle hosting numerous lakes and rivers that shaped the Noachian landscape. Whereas Earth evolved plate tectonics, Mars preserved two contrasting landscapes; the rough southern highlands and the smooth northern lowlands. How this 'Mars dichotomy' formed remains unanswered.

Evidence for Mars' aqueous past is recorded in surface minerals. Carbonates expand our understanding of carbon- and water cycles, as well as planetary habitability. They are commonplace on Earth but less abundant on the martian surface even though martian conditions could have been warm and wet; favourable for the formation of carbonates (in neutral pH water bodies). Although most carbonate on Mars formed under aqueous conditions (similar to Earth), some carbonates on Mars may have formed via other processes (that do not occur on Earth). Atmospheric $CO_2$ photochemistry, or energetic electrons discharging martian dust particles are examples. Sulphates on Mars are mostly hydrated, supporting their formation under water-rich conditions, mostly during the Hesperian (∼3.0–3.7 Ga ago). Hematite, clays, and amorphous hydrated silicates are more widespread, providing further evidence of an aqueous past.

As for martian water, studies of martian methane are motivated by the question of life on Mars. Whereas the mechanisms of biogenic methane formation on Earth have been identified, mechanisms of methane formation on Mars have yet to be elucidated. Methane levels are known to vary on Mars, but its origin remains unclear. Seasonal variation in martian methane can be attributed to either a biological origin or inorganic formation, e.g. via serpentinization, photochemistry or the reduction of $CO_2$ over acidic minerals and clays.

We are yet to ascertain the existence of past or present life on Mars. Habitable environments need to be studied, although a habitable environment does/did not necessarily harbour life. This said, the identificiation of biosignatures on Mars (whether textural, chemical, or isotopic) would confirm the existence of once-habitable environments. However, for many kinds of biosignature, there is no simple way to affirm that ancient life once existed in a specific extraterrestrial location. Microorganisms from Earth might not survive the harsh conditions on Mars surface, and no biomolecules would remain intact without the protection of minerals. However, Mars is not necessarily universally biocidal, and pigmented microorganisms are the most resistant to EUV exposure. The search for biomolecules should be focused in the subsurface of Mars.

A better understanding of martian processes requires the characterization of Mars across diverse spatial and temporal scales. Martian meteorites have been the only martian samples available for study at the microscopic scale. After being recognized as martian only decades ago, we are now exploring where on Mars these meteorites could have been ejected from. Cosmic ray exposure ages, determined via analyses of noble gasses, can indicate ejection times, grouping the martian meteorite ejection events and constraining the number of impact craters associated with them. An impact event that led to the Mojave Crater has been hypothesized to have ejected some of the shergottite martian meteorites. However, determination of martian meteorite source locations remains a challenge even with accurate mineralogical data on both Mars and its meteorites.

Mars exploration will intensify during the coming decade. The increasing number of missions to Mars will improve our understanding of the Red Planet, increasing the likelihood of human habitability. In parallel, advancements made in human space exploration on the ISS and eventually the Moon will catalyse human exploration of Mars. First and foremost, will be to determine where humans could safely inhabit Mars. The elevation of selected sites is important due to the landing risks; atmospheric breaking technologies need to be established. The major technology drivers required to achieve this include resource utilization, i.e. converting $CO_2$ into methane and oxygen for propellant fuel, life- and resource-support technologies, and water extraction from the polar ice and permafrost. Martian regolith and dust can be used with manufacturing technologies to produce human habitations. When sustainable agriculture is achieved, human habitability is finally supported. Further to that, additional risks associated with societal structures and human settlements require quantification and evaluation.

Humanity's encounter with a possibly inhabited world requires careful consideration of the United Nations resolutions on planetary protection. Contamination of Mars could occur from Earth as well as any potential bio-risk to Earth upon human or robotic return from Mars. This is particularly a concern towards the subsurface of Mars that could retain Earth-based organisms if contaminated. Precise countermeasures are required to prevent contamination from vehicles and instruments sent to Mars, and eventually from humans. An accurate characterization of the local martian environment for extant life is necessary prior to any human activities. If any forms of extant life were to be discovered, would it be ethical to tamper with their course of their evolution? Solving such ethical questions whilst sustaining human development on Mars is just as challenging as the technological feat itself.

Current and future Mars missions outline the technological achievements through which economic benefits arise and nations are brought together. Mars exploration will also support STEM education and careers addressing the global challenges we face together on Earth. Mars will ultimately act as a testing ground for humanity becoming an interplanetary species. This frontier is becoming more realistic, despite the technological challenges that need to be solved. These challenges, which involve the conservative and efficient use of resources, could also facilitate improved sustainability and economic prosperity on Earth.

**Acknowledgements.** HGC was supported by the NSFC 'Young International Scientist Award' Grant No. ED was supported by the Canadian Space Agency Grant No. 18FASASB13. 41750110488. MF acknowledges grants nos. 19-03314S of the Czech Science Foundation and ERDF/ESF 'Centre of Advanced Applied Sciences' (No. CZ.02.1.01/0.0/0.0/16_019/0000778). FXH was funded by the B-type Strategic Priority Program of the Chinese Academy of Sciences (Grant No. XDB41000000). Lauren Edgar of the United States Geological Survey is thanked for her contribution in the 'Robotic and remote exploration of Mars' section of this article. The National Space Science Centre of the Chinese Academy of Sciences and the Space Studies Board of US National Academy of Sciences' National Research Council's 9[th] and 10[th] New Leaders in Space forum organizers Dr Quanlin Fan, Chinese Academy of Sciences, China and Dr David Smith, US National Academy of Sciences, USA are also thanked for bringing together some Mars scientists for this article.

International Journal of Astrobiology                                                                                                                                             25